\documentclass[letterpaper]{aa}  

\usepackage{graphicx}
\usepackage{txfonts}
 \usepackage[figuresright]{rotating}
\usepackage{natbib}
\usepackage{supertabular}
\bibpunct{(}{)}{;}{a}{}{,}%
\begin{document}
   \title{Using VO tools to investigate distant radio starbursts
   hosting obscured AGN in the HDF(N) region } \subtitle{}
   \author{A.~M.~S.~Richards\inst{1}, T.~W.~B.~Muxlow\inst{1},
   R.~Beswick\inst{1}, M.~G.~Allen\inst{2}, K.~Benson\inst{3}, R.~C.~Dickson\inst{1},
   M.~A.~Garrett\inst{4}, S.~T.~Garrington\inst{1},
   E.~Gonzalez-Solarez\inst{5}, P.~A.~Harrison\inst{6}, A.~J.~Holloway\inst{1},
   M.~M.~Kettenis\inst{4}, 
R.~A.~Laing\inst{6}, 
E.~A.~Richards\inst{7},
H.~Thrall\inst{1}, H.~J.~van~Langevelde\inst{4,8},
   N.~A.~Walton\inst{5}, P.~N.~Wilkinson\inst{1} \and
   N.~Winstanley\inst{1}. }

   \offprints{A. M. S. Richards \\\email{amsr@jb.man.ac.uk}}

   \institute {Jodrell Bank Observatory, University of Manchester,
         SK11 9DL, Macclesfield, UK.
         \and Centre de Donn\'{e}es astronomiques de Strasbourg (UMR
         7550), F-67000, Strasbourg, France.
\and Mullard Space Science Laboratory, UCL,  Holmbury St. Mary, Dorking, Surrey, RH5 6NT, UK.
\and Joint Institute for VLBI in Europe, Postbus 2, 7990 AA Dwingeloo, The Netherlands.
\and Institute of Astronomy, Madingley Road, Cambridge, CB3 0HA, UK.
\and European Southern Observatory, D-85748 Garching bei M\"{u}nchen,
         Germany.
\and Department of Physics, Talledega College, Talledega, Alabama 35160, USA.
\and   Sterrewacht Leiden, Leiden University, Postbus 9513, 2300 RA Leiden, The Netherlands.       }

   \date{Received August 1, 2005; accepted March 16, 3005}

\abstract{A 10-arcmin region around the Hubble Deep Field (North)
  contains 92 radio sources brighter than $40$~$\mu$Jy which are
  well-resolved by MERLIN+VLA at 0\farcs2-2\arcsec\/ resolution
  (average size $\sim1$\arcsec). 55 of these have \emph{Chandra} X-ray
  counterparts in the 2-Ms CDF(N) field including at least 17 with a
  hard X-ray photon index and high luminosity characteristic of a
  type-II (obscured) AGN.
More than 70\% of the radio  sources have been classified as
  starbursts or AGN
  using radio morphologies, spectral indices and comparisons with
  optical appearance and rest-frame MIR emission. On this basis, starbursts
  outnumber radio AGN 3:1. } {We investigate the possibility that very
  luminous radio and X-ray emission originates from different
  phenomena in the same high-redshift galaxies.}  { This study extends
  the Virtual Observatory (VO) methods previously used to identify
  X-ray-selected obscured type-II AGN, to examine the relationship
  between radio and X-ray emission. We describe a VO cut-out server
  for MERLIN+VLA 1.4-GHz radio images in the HDF(N) region.
  }
  {The high-redshift starbursts have typical sizes of 5--10 kpc and
  star formation rates of $\sim1000$ M$_{\odot}$ yr$^{-1}$, an order
  of magnitude more extended and intense than in the local universe. There is no obvious correlation between radio and X-ray
  luminosities nor spectral indices at $z\ga1.3$. About 70\%
  of both the radio-selected AGN and the starburst samples were detected by
  \emph{Chandra}.
The X-ray luminosity indicates the presence of an AGN in at least half
  of the 45 cross-matched radio starbursts. Eleven of these are type-II
  AGN, of which 7 are at $z\ge1.5$. 
This distribution overlaps closely with the X-ray detected radio
  sources which were also detected by SCUBA.  In contrast, all but one
  of the AGN-dominated radio sources are at $z<1.5$, including the 4
  which are also X-ray selected type-II AGN.  The stacked 1.4-GHz emission at
  the positions of radio-faint X-ray sources is correlated with X-ray
  hardness.}  { Almost all extended
  radio starbursts at $z>1.3$ host X-ray selected obscured AGN. The
  radio emission from most of these ultra-luminous objects is
  dominated by star formation although the highest redshift
  ($z=4.424$) source has a substantial AGN contribution.
  Star-formation appears to contribute less than 1/3 of their X-ray
  luminosity. Our results support the inferences from SCUBA and IR
  data, that at $z\ga1.5$, star formation is observably more extended
  and more copious, it is closely linked to AGN activity and it is
  triggered differently, compared with  star formation at lower
  redshifts. }

   
\authorrunning{Richards et al.}
\titlerunning{Distant radio starbursts hosting obscured AGN}
   \maketitle
%

\section{Introduction}
\label{intro}

There is now general agreement that the number of vigorous
star-forming galaxies, and the star formation rate (SFR) within these
galaxies, increases dramatically at $z > 1$.  The details of how these
starburst galaxies relate to the high redshift Active Galactic Nucleus
(AGN) population are less clear.  Objects detected individually at
$z>1$ in radio and X-rays, by even the deepest available
exposures, are inevitably abnormally luminous.  Is it equally
inevitable that, in a galaxy detected in both regimes, all such bright
emission emanates from the same phenomenon, or can we separate
contributions from AGN and from starbursts if these coexist? 
 
The unprecedentedly sensitive observations of the Hubble Deep Field
(North) (HDF(N)\footnote{We use the term HDF(N) to describe
observations made in and around the original HDF; these form part of
the multi-wavelength GOODS (Great Observatories Origins Deep Surveys)
project \citep{Giavalisco04}.})  which commenced in 1996 provided the
first detailed attempts to quantify the star formation history of the
universe \citep{Madau96}.  The radio luminosity function evolves
rapidly with redshift, as $(1+z)^3$ for $0.5<z<1.5$ \citep{Cowie04a}.
Subsamples classified using optical spectra and X-ray power suggest
that the AGN luminosity function is declining at $z>0.9$ compared to
lower redshifts, while the reverse is the case for star-forming
galaxies.  This is supported by \emph{Spitzer} detections of
Ultra-Luminous IR Galaxies (ULIRGs) with IR luminosities $>10^{12}$
M$_{\odot}$. These show that the co-moving density of ULIRGs (with a
typical SFR of 200--300 M$_{\odot}$ yr$^{-1}$) at $z \sim 2$ was at
least 3 orders of magnitude greater than in the local universe
\citep{Daddi05}.

 Star-formation rates (SFR) measured from optical data only can be greatly
underestimated or overlooked altogether \citep{Reddy04}.  For example,
\citet{Cowie04a}, using optical spectra, were only able to classify
53\%  of the radio sources $<100$~$\mu$Jy in the HDF(N), finding 
that 28\% are star forming galaxies.
In contrast, over 2/3 of the 58
resolved sources  $<100$~$\mu$Jy in the HDF(N) were classified using
the radio-based criteria of \citet{Muxlow05}, containing 60\% starbursts.
Similarly, up to 90\% of the distant or obscured AGN revealed by deep
X-ray observations may be missed by optical surveys \citep{Bauer04}.

 Classification based on IR, sub-mm and radio properties is favoured
because local starburst galaxies show a strong peak in their spectral
energy distributions (SED) around 3~THz (100~$\mu$m)
\citep[e.g.][]{Yun02} which can be used to estimate the SFR
\citep{Condon92, Cram98, Yun01}.  The most striking evidence for
extraordinary levels of high-redshift star formation came from Sub-mm
Common User Bolometer Array (SCUBA) observations (\citealt{Hughes98};
\citealt{Smail02}; review by \citealt{Blain02}).  The median redshift
for SCUBA sources (SMG) in the HDF(N) with optical counterparts is at
least 2.
SMG have a typical SFR of
1000--2000 M$_{\odot}$ yr$^{-1}$, an order of magnitude greater than
in the most active local ULIRGs such as
Arp 220 (SFR 50--150 M$_{\sun}$ yr$^{-1}$).
 The FIR intensity is well-correlated with radio emission (on scales
greater than a few tens of pc)  \citep{Condon92,
Yun01}. \citet{Elbaz02}, \citet{Garrett02} and \citet{Chapman05}
have shown that the relationship is valid out to at least $z\sim 3$.

Star formation dominates the rest-frame MIR and FIR output even if an AGN is
present \citep{Downes98, Frayer98}
as emission due to dust heating by AGN declines steeply
from the NIR to the FIR \citep[e.g. Markarian 231,][]{Soifer00}.  The
observed ratio of X-ray to rest frame FIR luminosity is $\la$10\% in
local active galaxies even when a strong AGN is present and lower
still at high redshifts, especially for starburst-dominated sources
\citep{Alexander03XIV, Alexander05a}.

Almost half of the optical spectra available for 2-Ms X-ray sources in
the HDF(N) indicate the presence of star formation in the same galaxy
\citep{Sadler02, Barger05}.
There is evidence that radio and
X-ray emission has a common origin in starforming galaxies at
relatively low redshifts \citep{Alexander02XI}.  
\citet{Bauer02} derive a relationship between radio and
X-ray luminosities for 102  emission-line galaxies at $z \le 1.3$ (of
which only 2 sources at $z>1$ were detected in both
radio and X-rays in the data then available):
\begin{equation}
\log L_{\mathrm{X}}=(0.935\pm0.073)\log L_{\mathrm{R}} +
(13.141\pm1.650)
\label{Bauer}
\end{equation}
 where $L_{\mathrm{X}}$ and $L_{\mathrm{R}}$ are the X-ray
and radio rest-frame luminosities in W and W Hz$^{-1}$.  This does not
seem to hold so well for  samples extending to
higher redshift; \citet{Barger07} find no correlation between radio and X-ray
luminosities for optically-classified star-forming galaxies in the
HDF(N) brighter than 60 $\mu$Jy.

In this paper, we
investigate whether the relationship holds at high redshift using
classifications independent of optical detections and we
explore the properties of radio counterparts to the obscured AGN
(type-II AGN) identified from their hard X-ray photon indices and high X-ray
luminosities by \citet{Padovani04}. 

Many investigations of high-redshift star formation deliberately
exclude AGN hosts. We do not need to do this because we  use
sub-arcsec resolution to distinguish between different energy
sources in the same galaxy, which may correspond to different
classifications in different wavelength regimes.  
The
whole field has only been well-resolved by the \emph{HST} and by
MERLIN+VLA at 1.4 GHz.   The extent of radio emission from high-redshift 
galaxies in the HDF(N) is typically 1\arcsec--2\arcsec.

 We present the first detailed comparision between the highest
sensitivity MERLIN+VLA and \emph{Chandra} data ever taken and the
\emph{HST ACS} images.  The data used in this paper are described in
more detail in Section~\ref{data}, followed by a summary of the
Virtual Observatory and RadioNet\footnote{{\tt
http://www.radionet-eu.org/}} software which has made these results
possible, in Section~\ref{vos}.  In Sections~\ref{sec:lum}
and~\ref{sec:class} we explain how we derive the radio and X-ray
luminosities\footnote{We assume an empty Friedmann universe ($\Omega_0=0$), for
 consistency with \cite{Padovani04} and take $H_0$ $=70$ km
 s$^{-1}$ Mpc$^{-1}$.}
and deduce the origins of the emission,
based primarily radio data for the radio
sources and X-ray data for X-ray sources.  Their relationships are
explored in Section~\ref{sec:relate} and we present evidence for the
presence of embedded type-II AGN in radio starbursts in
Section~\ref{sbagn}.  We demonstrate statistically the presence of
faint radio emission associated with the majority of X-ray sources in
Section~\ref{radio-faint} and summarise our conclusions in
Section~\ref{conclusions}.


\section{HDF(N) Data, Cross-Identifications and Redshifts}
\label{data}

In this Section we introduce the radio observations and describe
briefly the X-ray and other data and tools used to make comparisons.
The positions, flux densities and spectral and photon indices of radio
sources with X-ray counterparts are listed in Table ~\ref{tab1}, along
with their redshifts and any IR or sub-mm detections.  All positions
given in this paper have been aligned with the VLA or MERLIN+VLA data
as these provide the most accurate reference frame, aligned with the
International Celestial Reference Frame (ICRF) to better than 15
milli-arcsec (mas) \citep{Muxlow05}.

\subsection{Radio observations of the HDF(N)}
\label{radio}
\citet{Muxlow05}, \citet{Richards00} and \citet{Richards98} describe
 the MERLIN and VLA observations of the HDF(N) made in 1996-7.  The
 VLA-only 1.4-GHz image contains 92 sources above its completeness
 limit of 40~$\mu$Jy per 2\arcsec\/ beam ($\sim5.5\sigma$) in a box of
 side 10\arcmin\/ (the 10-arcmin field), the `radio-bright' sample.
 The MERLIN field was centred on Right
 Ascension~12$^{\mathrm{h}}$~36$^{\mathrm{m}}$~49\fs4000,
 Declination\/ +62\degr~12\arcmin~58\farcs000 (J2000), hereafter taken
 as the reference position. 

\subsubsection{1.4 GHz MERLIN+VLA observations and other radio data}
\label{1.4}

 The combined MERLIN+VLA 1.4 GHz data reach an rms noise level of
 $1\sigma\sim3.5$~$\mu$Jy at $\la5$\arcmin\/ from the pointing centre,
 twice the sensitivity of the VLA-only data. Both arrays observed in
 wide-field mode, using short integration times and multiple narrow
 frequency channels across the bandpass in order to ensure that
 time-averaging and chromatic aberrations were less significant than
 the fundamental limitations of the primary beams. This is described
 in detail by \citet{Richards00} (his section 3.2 and figure 3) and
 \citet{Muxlow05} (their section 2). Computational limitations meant
 that the calibrated MERLIN and VLA data were separately Fourier
 transformed into multiple small dirty maps covering the region to be
 imaged; each pair was then combined and {\sc clean}ed.  Tests showed
 that, for an image with the same weighting and {\sc clean}ing, there
 was no appreciable difference between this method and data
 combination in the visibility plane \citep[][their figure
 1]{Muxlow05}.  The final combined images show $\la6$\% loss of flux
 at 5' from the pointing centre and there is no systematic radial
 distortion of the source contours \citep[][their figure
 C1]{Muxlow05}.

\citet{Muxlow05} resolved all 92 radio-bright sources at
 0\farcs2--2\arcsec\/ resolution, see Table~\ref{tab1},
 Fig.~\ref{las_sep.eps} and \citet{Muxlow05}. J123644+621133 is an
 FR~1 \citep{Fanaroff74} radio galaxy with jets extending over
 12\arcsec. Excluding this source, the mean angular size of sources at
 $z\la2$ is 1\farcs3, corresponding to $\sim10$ kpc at $z>0.8$.
Sources at $2\la
z\la3$ have a mean size of 8 kpc and the source at $z=4.424$ has a
size of 2 kpc.  The smaller apparent size of higher redshift sources
is probably at least partly due to the non-detection of fainter
extended emission (as well as being affected by the adopted cosmology)
and is not obviously linked to the inverse relationship between
angular size and redshift established for bright radio galaxies by
\citet{Barthel88}.

 VLA observations at 8.4 GHz covered the inner HDF(N) to a radius of
$\sim4$\arcmin\/ \citep{Richards98, Fomalont02} at a resolution of 3\farcs5,  
finding a total of 50 sources within the 10-arcmin field. 27 of these
sources were detected at $>40$~$\mu$Jy by MERLIN+VLA at 1.4 GHz. The
remainder cannot be classified using their radio morphologies and are
omitted from our analysis, apart from 7 which do have X-ray
counterparts, (Section~\ref{x}).  We refer to these as 8.4-GHz
selected sources. Their properties are given in Table~\ref{tab1},
including 1.4-GHz flux densities taken from \citet{Richards98}
where available or calculated using the spectral indices
described in Section~\ref{spindex}, so that the rest-frame
luminosities of the whole sample can be derived consistently in
Section~\ref{sec:lrad}.

We use $S_{\mathrm{R}}$ to denote the total radio flux density
measured by the VLA at either frequency, further subscripted by the
specific frequency only where relevant.

 Four of the 92 sources were detected at (4 -- 20)-mas resolution by
the EVN (European VLBI Network) and global VLBI \citep{Garrett01,
Chi06}.
 At the other extreme, the Westerbork Synthesis Radio Telescope at 15\arcsec\/
 resolution detected $\sim10$\% more sources than the VLA
 \citep{Garrett00}.  Further VLA images on larger scales 
  are in preparation \citep{Morrison06} and  recent
 low-frequency observations have been made using the GMRT (Lal,
 D. V., in prep).

\subsubsection{Radio spectral indices}
\label{spindex}

The radio
spectral index $\alpha$ is given by 
\begin{equation}
S_{\mathrm{R}}\propto\nu^{-\alpha}.
\label{alpha}
\end{equation}  
Asymmetric uncertainties are represented by e.g.
$\alpha^{+\sigma_{\alpha_+}}_{-\sigma_{\alpha_-}}$. \citet{Richards00}
and \citet{Richards98} provide spectral indices and uncertainties for
sources detected by the VLA at both frequencies. 
If
sources within the overlapping spatial region were detected at 1.4
(8.4) GHz only, $\alpha$ is a lower (upper) limit depending on the
local noise in the 8.4 (1.4) GHz field.  In the case of a 1.4-GHz-only
detection $\sigma_{\alpha_-}$ is approximated as twice the error due to
the 1.4 GHz detection,
\begin{equation}
\sigma_{\alpha_-} = \frac{2}{\ln[8.4/1.4]} \frac{\sigma_{S_{\mathrm{R1.4}}}}{S_{\mathrm{R1.4}}}
\end{equation} 
and $\sigma_{\alpha_+}$ is given by an analogous expression for
8.4-GHz-only detections.  

Radio sources classified as AGN or as starbursts (see
Section~\ref{sec:class}) detected at both frequencies had spectral
indices in the ranges ($-0.4<\alpha<1.4$) and ($0.3<\alpha<1.7$)
respectively; all unclassified sources had $\alpha$ within the extrema
of these ranges.  Where $\alpha$ is a lower limit we set
$\sigma_{\alpha_+}$ to the relevant upper limit. e.g. ($1.7-\alpha$)
for starburst or unclassified sources.  The errors in $\alpha$ for
8.4-GHz selected sources were deduced in a similar fashion for the
opposite limits.  For sources outside the 8.4 GHz field we adopted
typical values of $\alpha$ of 0 and 0.8 for AGN and starbursts
respectively and an average of 0.4 for unclassified sources, using the
extrema to deduce the uncertainties, so that for example a starburst
would have $\alpha=0.8\pm^{0.9}_{0.5}$.

\subsection{X-ray data}
\label{x}

The \emph{Chandra} X-ray observatory made a total of 2 Ms multi-band
exposures of the HDF(N) \citep{Alexander03XIII}.  
All X-ray flux densities, counts and luminosities given in this paper
refer to the \emph{Chandra} full band from 0.5--8.0 keV unless
otherwise stated.  Soft-band values are used for J123709+620841 and
J123646+621445 as they were not detected in the full band.
 There are 100 sources in common within the area of overlap between
the whole VLA and \emph{Chandra} fields of view, with a median offset
of $\sim$0\farcs2 after small corrections to align the X-ray frame
\citep{Alexander03XIII}. 

\subsubsection{X-ray -- radio counterparts}

The \emph{Chandra} observations completely enclose the radio 10-arcmin
field and the decline in sensitivity in both images towards the edges
of this region is less than 6\%.  This field contains 253 X-ray
sources with position uncertainties 0\farcs3--0\farcs9.  Fifty-five
(60\%) of the radio-bright sources have X-ray counterparts within
0\farcs9 of the radio peak; the separation is $<$0\farcs4 for 42 of
these.  Increasing the cross-match search radius up to 2\arcsec\/
failed to produce any more matches. One or two additional matches
appear for each additional arcsec radius from 2--5\arcsec\/. Each of these
radio sources also has a counterpart at $<$0\farcs9; in about half
these cases the multiple associations appear to be genuine
(e.g. similar redshifts).  We consider that we can only be confident that the
emission is coming from the same galaxy for the 55 unambiguous matches
at $<$0\farcs9 separation. These make up 22\% of the X-ray detections
in the 10-arcmin field. Seven additional 8.4-GHz selected sources have
X-ray counterparts within their combined position uncertainties.

\begin{figure}
\centering
\includegraphics[angle=0,width=8.8cm]{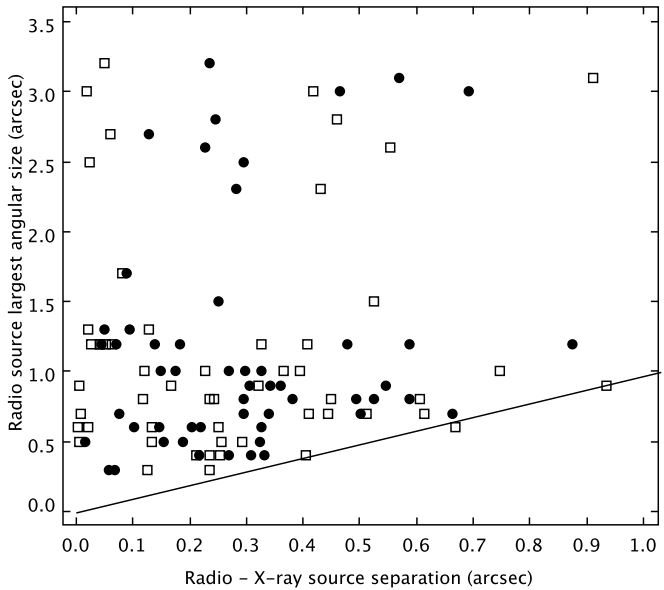}
   \caption{ The filled symbols show the measured angular sizes of
           radio sources with X-ray counterparts compared with the
           radio-X-ray peak separation.  The sloping line has a
           gradient of unity, showing that all the radio sources have
           an angular size greater than the distance to their X-ray
           counterpart. The hollow symbols show the measured angular
           size as a function of randomised X-ray position errors (see
           text).  }
      \label{las_sep.eps}
\end{figure}

We compared the largest angular size of each radio-bright source with
the X-ray -- radio source separation, represented by the solid circles
in Fig.~\ref{las_sep.eps} (J123644+621133, with an angular size of
12\arcsec, has been omitted).  In every case the X-ray peak is no
further from the radio peak than the most extended radio emission.  We
produced randomised X-ray position errors, in a Gaussian distribution,
such that 80-90\% of the X-ray positions were within the published
errors of 0\farcs3--0\farcs9 \citep{Alexander03XIII}, which are plotted
as hollow squares. The radio peak position errors are negligible in
comparison ($\la$0\farcs1). There is no evidence for any systematic
excess in the measured source separations with respect to the X-ray
position errors but peak offsets of $\la1$\arcsec\/ cannot be ruled
out.

\subsubsection{X-ray photon indices}

The X-ray photon index $\Gamma$, for flux density  $F_{\mathrm{X}}$ in
$10^{-18}$ W m$^{-2}$ at energy $E$ keV is defined by
\begin{equation}
 F_{\mathrm{X}}(E)\propto E^{-\Gamma}
\label{gamma}
\end{equation}
and $\Gamma$ is equivalent to $\alpha+1$ (where $\alpha$ is the
spectral index).  All values measured by
\citet{Alexander03XIII} lie in the range $-1\la\Gamma\la2$, who give
upper and lower bounds to the uncertainty, $\sigma_{\Gamma_+}$ and
$\sigma_{\Gamma_-}$ only for sources detected in more than one
sub-band.  If a source is only detected in the full and soft bands
then $\Gamma$ is a lower limit so we take $\sigma_{\Gamma_+}=2-\Gamma$
and approximate $\sigma_{\Gamma_-}$ as twice the error due to the
uncertainty in the soft band counts. If a source is only detected in
the full and hard bands then $\Gamma$ is an upper limit so we take
$\sigma_{\Gamma_-}=\Gamma - (-1)$ and approximate $\sigma_{\Gamma_-}$ as
twice the error due to the uncertainty in the hard band counts.  If
sources were only detected in the full band, \citet{Alexander03XIII}
give an estimated value of $\Gamma =1.4$ and we adopt
$\sigma_{\Gamma_-}= \sigma_{\Gamma_+}= 0.6$.  In all cases we constrain
$\sigma_{\Gamma_-} \le 0.6$ in order to avoid non-physical negative
limits on $L_{\mathrm{X}}$ (see Section~\ref{sec:lx}).

\subsection{Hubble Space Telescope observations}
The original HDF and surrounding fields (out to a distance of
$\sim5$\arcmin) was observed by the \emph{HST WFPC2} in 1996
\citep{Williams96}.  In 2003 the GOODS project used the \emph{HST ACS}
to re-observe the HDF region in the F435W, F606W, F775W and F850LP
filters (\emph{B}, \emph{V}, \emph{i} and \emph{z} bands).  We find
that the GOODS images and source catalogue r1.1z \citep{Giavalisco04}
require a linear shift of --0\farcs342 in Declination to align them
with the ICRF.

\subsection{IR and sub-mm sources}
\label{sec:FIR}
 The \emph{ISO} fields and the \emph{Spitzer} catalogue published by
\citet{Teplitz05} only cover part of the 10-arcmin field so it is only
possible to give meaningful statistics for the fractions of the IR
catalogues detected at other wavelengths (not visa versa).  More
quantitative analysis will be available using further \emph{Spitzer}
results at 24 $\mu$m (see e.g. \citealt{Beswick06}).  Extensive SCUBA
searches have been made over most of the HDF(N).

\subsubsection{Mid-infrared detections}
\label{IR}

One hundred sources were detected in the inner HDF(N) by \emph{ISO} at
  7 or 15 $\mu$m \citep{Aussel99}. Although the beam size was
  3--6\arcsec\/ the tight correlation between radio and 15~$\mu$m flux
  densities out to at least $z=3$ (\citealt{Garrett02},
  \citealt{Elbaz02}, see Section~\ref{intro}) supports the association
  of radio and IR sources within the position errors even if they
  overlap more than one optical source. 28 radio-bright sources lie
  within   the \emph{ISO} field, of which 17 have \emph{ISO} counterparts
  \citep{Muxlow05}. All matched sources were detected at 15-$\mu$m
  except for J123656+621301. This is nonetheless an extended diffuse
  radio source with a very steep spectrum characteristic of a
  starburst.  An additional 7-$\mu$m source in the catalogue of
  \citet{Goldschmidt97} is matched with the FR~1 J12364+621133.

 The radio-MIR association has been reinforced by recently-published
  \emph{Spitzer} observations at 16~$\mu$m \citep{Teplitz05}. 18
  \emph{Spitzer} sources have MERLIN+VLA counterparts within
  1\farcs2. Half of these lie outside the \emph{ISO} fields. Of the
  other nine, 7 already had \emph{ISO} counterparts (including the
  very red source J123651+621221 at $z=2.71$; \citet{Teplitz05}
  associate the IR emission with an elliptical galaxy at a similar
  separation but lower redshift).  The other two, J123633+621005 and
  J123708+621056, lie close to the edges of the \emph{ISO} field where
  its noise was higher.  We cannot confidently associate
  J123646+621445 with the \emph{Spitzer} source 1\farcs7 to the SW as
  they have two separate optical counterparts. An \emph{ISO} source
  lies within 3\arcsec\/ of the 16~$\mu$m source but further from
  J123646+621445. There are no further candidate radio-IR matches
  within 2\arcsec.

The combined \emph{Spitzer} and \emph{ISO} data contain 205 separate
15- or 16-$\mu$m sources within the 10-arcmin field of which a quarter
(53) have X-ray counterparts.  Even fewer (26, 13\%) have radio-bright
counterparts, but almost all of these (21/26) are also X-ray
detections.  This complements the tendency, noted by
\citet{Alexander02XI}, that optically identified (emission line)
15~$\mu$m starbursts with X-ray emission are more likely to have radio
counterparts than those without.  Four of the 7 8.4-GHz selected
sources with X-ray counterparts have \emph{ISO} counterparts, 3 of
which were also detected by \emph{Spitzer}.

\subsubsection{SCUBA detections}
\label{scuba}

Several sets of observing and data reduction techniques have produced
various SCUBA catalogues optimised for different regions and
properties \citep[e.g.][]{Serjeant03, Borys04, Wang04, Chapman05}.
The techniques used to minimise ambiguity in cross-identifications are
summarised in \citet{Muxlow05}.  The most comprehensive list is
currently provided by \citet{Borys04} (the revisions by \citet{Pope05}
do not affect any radio-bright sources).  We use all their secure
identifications between SMGs and radio-bright sources.  We also
include the additional identifications of J123622+621629 and
J123711+621325 made by \citet{Chapman05}. We do not include the SMGs
known as HDF 850-1 and 850-6 as most authors conclude that they do not
have radio-bright counterparts.  
J123608+621431 is
$\sim$3\arcsec\/ from the nearest X-ray source so it is not included
in the detailed analysis in this paper, but both objects are within
the larger error circle of a SCUBA source.  We reject the
identification of J123646+621445 for reasons similar to those given in
Section~\ref{IR} with respect to IR sources.

This leaves 16 radio-bright sources in the 10-arcmin field with SCUBA
counterparts, of which 11 were also detected by \emph{Chandra}; one
further 8.4-GHz source has both SCUBA and X-ray detections.  All these
sources have either spectroscopic or photometric redshifts, which we
adopt in preference to redshifts derived from the 1.4-GHz/850-$\mu$m
flux density ratio in order to avoid circular arguments.

\subsection{Redshifts}
\label{z}
61 radio-bright sources and 140  \emph{Chandra}
sources in the 10-arcmin field have measured redshifts, including 50
of the 55 radio-bright X-ray sources.  19 of the 8.4-GHz selected
sources also have measured redshifts, including 7 with X-ray
counterparts.

Table~\ref{tab1} gives our adopted redshift measurements,
uncertainties $\sigma_{\mathrm{z}}$ and references for the sources
detected in both regimes.  We include the published errors,
$\sigma_z$, where given. If not we adopt $\sigma_z=0.003$ for
spectroscopic redshifts, which were all obtained using the Keck LRIS or
instruments with resolution as good or better.  The uncertainties
in photometric redshifts are $\le z/4$ apart from J123725+621128 where
$1<z<2$ was estimated from the \emph{K}:\emph{z} band flux density
ratio \citep{Hornschemeier01}.  The uncertainties do not include
possible misidentifications of objects or of spectral lines, nor
instabilities in photometric fitting.
In most cases the differences between different redshift
estimates for the same source are small, or have been discussed and
resolved in the literature. The redshift for J123616+621513 has now
been revised to 2.58 \citep{Chapman04b}.
We adopt recently-published redshifts for  faint \emph{NICMOS} or
\emph{ACS} galaxies associated with
the radio sources J123606+621021, J123642+62133, J123651+621221 and
J123716+621512, in preference to the photometric redshifts derived by
\citet{Barger03} for their X-ray counterparts using more widely
separated, older optical detections.

\begin{figure}
\centering
\includegraphics[angle=0,width=8.8cm]{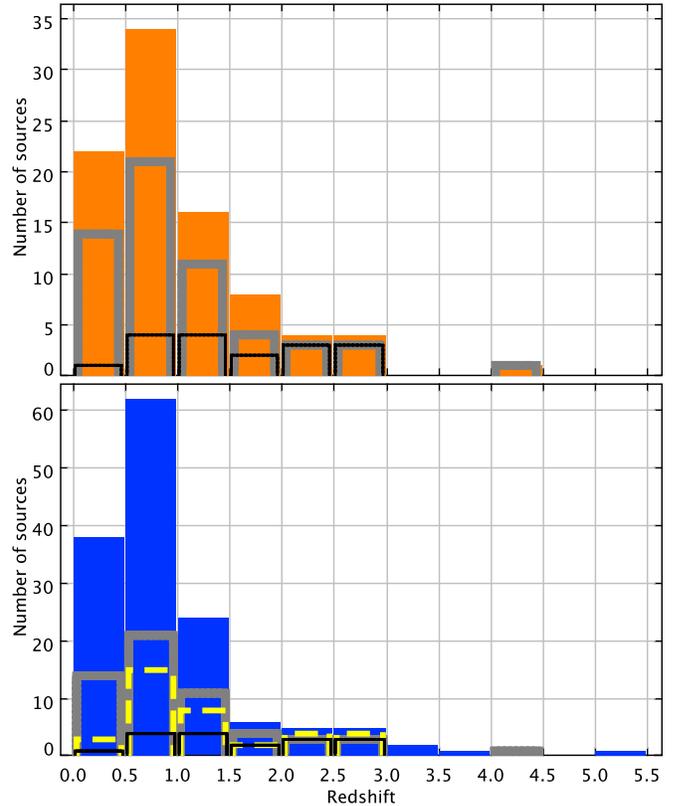}
   \caption{The distribution of sources with published redshifts in
   the 10-arcmin field. The filled  blue area in the lower panel
   shows X-ray sources and the dashed yellow line shows X-ray selected
   type-II AGN.  The filled orange area in the upper panel shows radio sources.
   In both panels the cross-matched X-ray and radio sources, and those
   which are also X-ray selected type-II AGN, are shown by the thick grey
   and thin black lines, respectively.  }
      \label{ZX_AGN2_XRAGN2.eps}
\end{figure}

The redshift distributions of radio and X-ray sources and of objects
detected in both regimes are compared in
Fig.~\ref{ZX_AGN2_XRAGN2.eps}.  All three distributions peak at
$0.5<z<1.0$ but the fraction of X-ray sources with measured redshifts
which are radio-loud changes from less than a third at $z<1$ to a half
or greater at higher $z$. A
similar increase in codetections with redshift is seen in the fraction
of radio sources which are X-ray-selected type-II AGN.  We used the
Kolmogorov-Smirnov test to investigate the relationship between the
redshift distributions of the radio and X-ray sources.  We found that
there is a 93\% probability that radio and X-ray sources at $z<1.1$
are drawn from the same population and a 98\% probability for sources
at $z>1.1$, but this drops to an insignificant probability of 27\% for
all redshifts considered together.  This only makes sense if the radio
counterparts to X-ray sources at lower redshifts are a separate
population from those at higher redshifts.  These implications are
discussed in Sections~\ref{xorig} and~\ref{sbagn}.

\section{New Technology}
\label{nt}

\subsection{Virtual Observatory facilities}
\label{vos}
We made use of a wide range of published surveys and catalogues.
These were obtained using the Vizier\footnote{{\tt
http://vizier.u-strasbg.fr/}} service where possible, in order to
select sources in the exact area covered by the radio data, and obtain
tables in VOTable\footnote{{\tt
http://cdsweb.u-strasbg.fr/doc/VOTable/}} format for ease of further
manipulation.  Other data (e.g. in IPAC format) were converted to
VOTable using TopCat\footnote{{\tt http://www.starlink.ac.uk/topcat}},
which preserves accuracy equivalent to full double precision.  Sources
were crossmatched using either the AstroGrid\footnote{{\tt
http://www.astrogrid.org}} Xmatch tool or TopCat, which allowed us to
identify and correct for any systematic linear offsets due to
astrometric errors.  We also used TopCat to calculate luminosities and
other derived quantities (Section~\ref{sec:lum}) presented in the
tables, and to prepare many of
the plots. 

\begin{figure*}
\sidecaption
\includegraphics[angle=0,width=12cm]{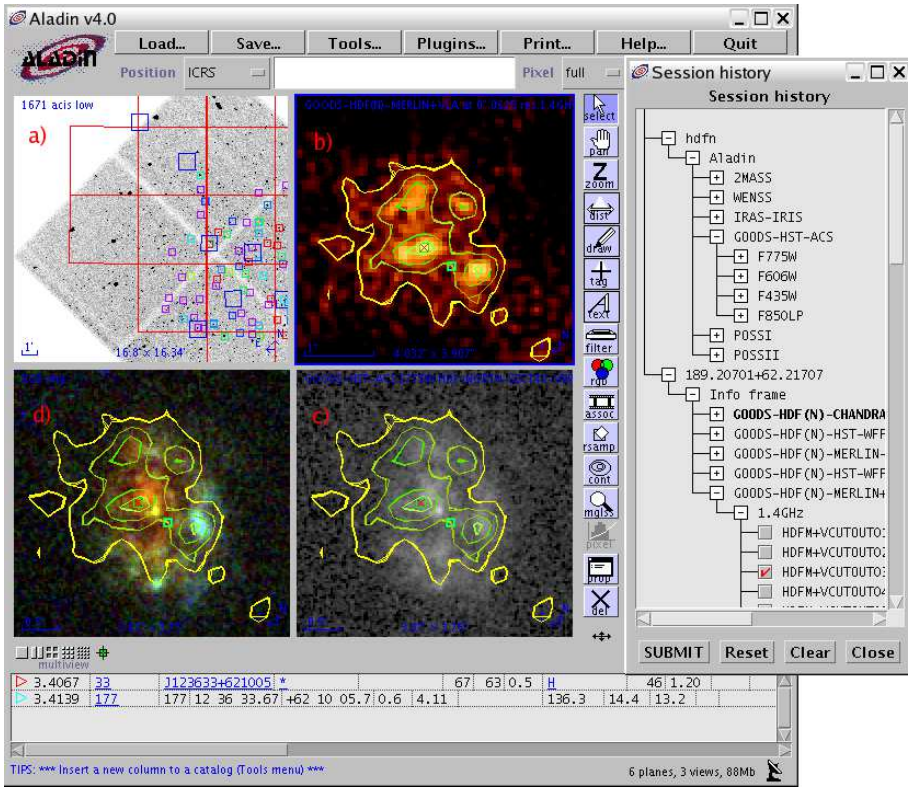}
   \caption{ The use of the Euro-VO Aladin and PLASTIC to investigate
   the starburst candidate J123633+621005.  Clockwise from top left,
   the panels show: a)~\emph{Chandra} image overlaid with the outlines
   of \emph{HST ACS} field boundaries; clicking in the appropriate
   square selects the relevant image(s) for loading.  The coloured
   symbols show radio sources, with symbol size proportional to source
   size and shade proportional to redshift. b)~Radio contours for
   J123633+621005; the red cross and blue square mark radio and X-ray peaks,
   respectively. c)~\emph{ACS} F775W image d)~\emph{ACS} false colour
   composite of F435W, F606W and F850LP bands, all overlaid with radio
   contours.  The
   scale bars in the bottom left of each panel represent a) 1 arcmin,
   b) 1 arcsec, c) and d) 0\farcs5. }
      \label{Aladin123633+621005.eps}
\end{figure*}

The original images have resolutions from $\sim$0\farcs015 ({\em
HST}) to several arcsec (\emph{ISO}, SCUBA).  The MERLIN+VLA and {\em
HST} maps are made up of many small panels each containing about a
million pixels.  We used the Aladin visualisation tool as modified for
the Euro-VO\footnote{{\tt http://www.euro-vo.org}} to find, cross-identify
and visualise regions of interest on such different scales; an example
is shown in Fig.~\ref{Aladin123633+621005.eps}.  The International
Virtual Observatory\footnote{{\tt http://www.ivoa.net}} Simple Image
Access Protocol (SIAP) for descriptions of images and their locations is used
to locate the corresponding fields in \emph{Chandra}, MERLIN+VLA and
\emph{HST} images despite the different resolutions, image sizes and
even orientations.  The PLASTIC\footnote{\tt http://plastic.sourceforge.net/} protocol developed for
VOTech\footnote{{\tt http://eurovotech.org/}} allows any of these VO
tools to manipulate the same data. 

\subsection{RadioNet tools and dynamic radio imaging}
\label{parsel}
It is now feasible to image much larger radio fields in entirety,
compared with the epoch when the MERLIN+VLA observations were made,
thanks to increased computing power, improved algorithms in {\sc aips}
and the use of Virtual Observatory standards and tools for data
management.  We made 81 slightly overlapping square images, each of
$1024\times$0\farcs0625 pixels on a side. We combined these images
into a single ($8\times8$)~arcmin$^2$ 1.4-GHz image, hereafter the
8-arcmin field. This contains over 67~Mpixels, covering most of the
maximum sensitivity regions of both the radio and X-ray images, with
good overlap with the \emph{ACS} data. This will be extended to cover
the 5-arcmin radius region of near-optimum radio sensitivity. We
describe our use of the new 8-arcmin HDF(N) image in
Section~\ref{radio-faint}.

The range of baseline lengths in the combined MERLIN+VLA data means
that maps can be extracted at resolutions of 0\farcs2--2\arcsec
depending on whether the observer wants to investigate potential
compact hot spots or faint extended emission.  \citet{Muxlow05} gives
a full description of the method which was used to produce the earlier
hand-processed images. We now provide an automatic imaging service
which extracts the required region and convolves it with the chosen
restoring beam within this resolution range. This uses the
python-based package {\tt ParselTongue}\footnote{{\tt http://www.radionet-eu.org/rnwiki/ParselTongue}} \citep{Kettenis06}, developed in the RadioNet Consortium, to provide a scripting interface between AstroGrid and
`classic' {\sc aips}. The AstroGrid workbench offers a simple dialogue
box for the user to select image size, resolution and region within
the HDF(N).  These parameters are passed to the MERLIN archive server
which uses {\tt ParselTongue} to extract the required image.  A
pointer to the image and a basic (SIAP-compliant) description is
either returned straight to the user or can be used to pass it to
another VO-enabled tool such as Aladin or a source extractor. This VO
tool and the complementary MERLINImager (which operates on the
visibility data for other MERLIN archive data) are the first to allow
an astronomer to obtain customised radio images without having to
install their own specialised radio data reduction package. Moreover,
only the required image (at most 0.25 GB) is moved over the internet
to the point of use; the parent data set, which can be many GB, is
processed in situ.

\section{Estimation of radio and X-ray luminosities}
\label{sec:lum}

There is no correlation between radio and X-ray flux densities for the
whole  cross-matched sample nor for any subsets; however this is not
surprising given the wide span of redshifts and the different
behaviour of the spectral/photon indices for different sources.  We
therefore compared the $K$-corrected luminosities for all 50 sources
with measured redshifts.

\subsection{Rest frame 1.4-GHz radio luminosity}
\label{sec:lrad}
Table~\ref{tab1} lists the observed-frame 1.4-GHz total radio flux density per
source, $S_{\mathrm{R}}$, in $\mu$Jy.  The flux densities and their
uncertainties ($\sigma_{S_{\mathrm{R}}}$) are given in \citet{Muxlow05} and
\cite{Richards98}. 

We assume an empty Friedmann universe ($\Omega_0=0$), for
 consistency with \cite{Padovani04} and take $H_0$ $=70$ km
 s$^{-1}$ Mpc$^{-1}$.  
The radio rest-frame luminosity $L_{\mathrm{R}}$, taking into account the {\em
K-}correction and the expansion of the bandwidth in the observed
frame, is given in W Hz$^{-1}$ by
\begin{equation}
L_{\mathrm{R}} = \frac{S_{\mathrm{R}}}{10^{32}} 4 \pi ({d}\times{3.086\times10^{22}})^2(1+z)^{\alpha-1}
\label{Lum1.4}
\end{equation}
where $d$ is the luminosity distance in kpc, given by
\begin{equation}
d = \frac{cz}{H_0}(1+z/2)
\label{dist}
\end{equation}
where the speed of light, $c$, is in km s$^{-1}$.
The lower bound on the uncertainty $\sigma_{L_{\mathrm{R}}-}$
is given by
\begin{equation}
\sigma_{L_{\mathrm{R}}-}=\sqrt{\sigma_{L_{\mathrm{R}}(\alpha-)}+\sigma_{L_{\mathrm{R}}(S_{\mathrm{R}})} + \sigma_{L_{\mathrm{R}}(z)}}
\label{lrerr}
\end{equation}
where the partial errors in $L_{\mathrm{R}}$ due to the lower bound on the spectral index
error, due to the flux density error and due to the redshift error are
given by:
\begin{eqnarray}
\label{Lrerra}  
\sigma_{L_{\mathrm{R}}(\alpha-)}&=&L_{\mathrm{R}} \sigma_{\alpha_-}\ln(1+z) \\
\label{Lrerrs}  
 \sigma_{L_{\mathrm{R}}(S_{\mathrm{R}})}&=&L_{\mathrm{R}} \frac{\sigma_{S{\mathrm{R}}}}{S_{\mathrm{R}}}\\
\sigma_{L_{\mathrm{R}}(z)}\;\;&=&L_{\mathrm{R}} \sigma_z \left(\frac{2(1+z)}{z(1+z/2)}+ \frac{\alpha-1}{1+z}\right).
\label{Lrerrz}  
\end{eqnarray}
Substituting $\sigma_{\alpha_+}$ for $\sigma_{\alpha_-}$ in
Equation~\ref{Lrerra} gives $\sigma_{L_{\mathrm R}(\alpha+)}$ which
then replaces $\sigma_{L_{\mathrm R}(\alpha-)}$ in
Equation~\ref{lrerr} to give the upper bound to the uncertainty in
$L_{\mathrm{R}}$, $\sigma_{L_{\mathrm{R}}+}$.  Table~\ref{tab2} gives
$L_{\mathrm{R}}$ and its upper and lower bounds
($L_{\mathrm{R}}-\sigma_{L_{\mathrm{R}}-}$ and
$L_{\mathrm{R}}+\sigma_{L_{\mathrm{R}}+}$).

\subsection{Rest-frame 0.5-8.0 keV  X-ray luminosity}
\label{sec:lx}
The rest-frame X-ray luminosity is given by the analogy of
Equation~\ref{Lum1.4}.
\begin{equation}
L_{\mathrm{X}} =  \frac{F_{\mathrm{X}}}{10^{18}} 4 \pi ({d}\times{3.086\times10^{22}})^2(1+z)^{\Gamma-2}
\label{LumX}
\end{equation}
The observed  flux densities are given in Table~\ref{tab1},
taken from \citet{Alexander03XIII}.  This does not give flux density
uncertainties so we assume that the relative uncertainty in the flux
density is equivalent to the relative uncertainty  in the counts
($C$).
$\sigma_{\mathrm{C-}}$ and $\sigma_{\mathrm{C+}}$ are the lower and
upper bounds on the uncertainties in the counts. 
The lower limit to the luminosity uncertainty $\sigma_{L_{\mathrm{X}}-}$ is given by the sum in quadrature of uncertainties due to the
lower limits to the errors in $\Gamma$ and $F_{\mathrm{X}}$ and to the
error in $z$
\begin{eqnarray}
\label{Lxerrg}  
\sigma_{L_{\mathrm{X}}(\Gamma-)}\,\,\,&=&L_{\mathrm{X}} \sigma_{\Gamma-}\ln(1+z) \\
\label{Lxerrf}  
 \sigma_{\mathrm{L_{\mathrm X}}(F_{\mathrm{X}}-)}&=&L_{\mathrm{X}} \frac{\sigma_{C-}}{C}\\
 \sigma_{\mathrm{L_{\mathrm X}}(z)}\,\,\,\,\,&=&L_{\mathrm{X}}\sigma_z\left(\frac{2(1+z)}{z(1+z/2)}+\frac{\Gamma-2}{1+z}\right)
\label{Lxerrz}  
\end{eqnarray}
and the upper limit to the uncertainty, $\sigma_{L_{\mathrm{X}}+}$, is given
by an analogous expression. Table~\ref{tab2} gives $L_{\mathrm{X}}$
and its upper and lower bounds ($L_{\mathrm{X}}-\sigma_{L_{\mathrm{X}}-}$ and $L_{\mathrm{X}}+\sigma_{L_{\mathrm{X}}+}$).

\section{Origins of Radio and X-ray emission}
\label{sec:class}
In this Section, we discuss diagnostics for the specific origins of
the radio and X-ray emission, based on the references and discussion
of Section~\ref{intro}, applied to the derived source properties. We
keep the initial radio source classification independent of X-ray
properties (and vice versa) as we wish to investigate whether the observed radio and
X-ray emission comes from different sources within the same
galaxies. In particular, we do not use published radio-X-ray
relationships such as Equation~\ref{Bauer} for classification, but
compare our results with this in the next Section,~\ref{sec:relate}.
\subsection{The nature of the radio emission}
\label{radorig}
 The main diagnostics for the origins of the radio-bright emission are:
\begin{enumerate}
\item {\bf Morphology} The presence of a bright, compact core or clear
  radio jets/lobes suggests the presence of a radio AGN.  Extended
  emission (kpc scales) with no obvious peaks or jets/lobes is more
  likely to originate from star formation.
\item {\bf Spectral Index} Radio emission with $\alpha\la0.4$ is likely to
  be powered by an AGN. Steeper-spectrum emission is usually of starburst
  origin but could indicate a lobe-dominated AGN system, hence the
  need to inspect the structure.
\item {\bf The Radio-FIR link} Rest-frame FIR or MIR emission bright
  enough to be detectable in the HDF(N) (including by SCUBA, for
  high-$z$ sources) indicates a ULIRG-like intensity or greater, very
  likely to be of starburst origin, found to be closely correlated
  with radio emission. Even when there is separate evidence for the
  presence of an AGN (see Section~\ref{intro}), rest-frame MIR
  emission at or greater than ULIRG-like luminosity is far in excess of
  the predictions of any known AGN model \citep[][ see also
  Section~\ref{scubasec}]{Alexander05a} and AGN are unlikely to
  be responsible for more than 20\% of the bolometric luminosity
  \citep{Alexander03XIV, Alexander05b}.
\item {\bf Optical appearance} If bright radio and optical galaxy
 cores coincide, this is further evidence for the presence of an AGN.
 Extended radio emission outside the apparent optical extent of a
 galaxy is likely to be of jet origin. On the other hand, extended
 radio emission within the optical galaxy, especially if associated
 with bright optical knots, indicates active star formation. Very
 disturbed or interacting galaxies with associated diffuse radio
 emission provide supporting evidence of conditions for starburst
 activity.  As explained in Section~\ref{intro}, obscuration can hide
 optical signatures, especially in starbursts, so absence of (optical)
 evidence is not conclusive evidence of absence of a particular radio
 emission mechanism.
\end{enumerate}
\citet{Muxlow05} described in detail the application of these criteria
in classifying radio sources as Starburst (SB), Active Galactic
Nucleus (AGN), or
unclassified (U). Note that the AGN status covers any emission powered
by an AGN, whether it arises from accretion or from jets/lobes. Using
the MERLIN+VLA data, we are able to distinguish between extended lobes
and cores, but not between pc-scale jets and the core itself, although
the resolved emission of AGN origin must be from jets or lobes.  The
presence of two unambiguous FR~1 in the HDF(N) already represents a
greater space density than would be expected from local number counts
\citep{Snellen01}, making it unlikely that a high proportion of the
unidentified extended or steep-spectrum sources have large radio
lobes.

VLBI results support the AGN interpretation of compact, flat-spectrum
radio cores.  Extremely compact radio cores (brightness temperature
($>10^5 - 10^6$) K) were detected in J123642+621331, J123644+621133,
J123646+621404 and J123652+121444 using the European VLBI Network
(EVN) and global VLBI \citep{Garrett01, Chi06}, confirming the
presence of an AGN. J123644+621133 is unmistakably an FR~1.  The EVN
recovers all the VLA flux from J123646+621404. About 1/3 of the VLA
flux from J123652+121444 is present in the 4-mas resolution global
VLBI image but the source is known to be variable \citep{Richards98}.
All three sources have flat or inverted spectra.  J123642+621133 is
discussed in more detail in Section~\ref{1331}; in summary we infer
that it consists of compact AGN-powered emission embedded in a more
diffuse starburst.  The MERLIN+VLA data suggest that this is also the
situation for J123635+621424 and J123642+121545.

Table~\ref{tab2} gives our classifications for the objects with X-ray
counterparts.  The recent \emph{ACS} and \emph{Spitzer} data and
improved SCUBA source lists have allowed us to strengthen the
classification of a number of sources.  We have changed the classification of
4 sources, as follows. J123622+621544 was tentatively assigned AGN
status by \citet{Muxlow05} 
but the \emph{ACS} image shows that the radio emission is extended
over bright optical knots in a distorted spiral, not seen in the
original CFHT plate (Canada-France-Hawaii Telescope,
\citealt{Barger99}). It is also a new MIR detection by \emph{Spitzer}
and has a radio spectral index $>0.6$ so we reclassify J123622+621544
as a starburst. We infer from the \emph{Spitzer} and \emph{ACS} images  that
two previously unclassified sources with steep radio spectra are
starbursts. J123629+621046 is extended, with a red optical
counterpart which is either a distorted galaxy with a dust lane or two
interacting galaxies.  J123641+620948 has a compact core but the \emph{ACS}
image confirms the suggestion by \citet{Cohen00} that it is associated
with two interacting spiral galaxies.
The \emph{ACS} morphology allows us to reclassify J123642+621545 as a
starburst candidate with a possible AGN core, as its extended radio
emission overlays blue knots in the arms of a face-on spiral.  It also
possesses a relatively bright compact radio and optical core.  It was
detected by \emph{ISO} and has an intermediate radio spectral index of
0.5.

Note that the classifications of radio-bright sources are made primarily on the
basis of radio properties such as morphology and/or spectral index
(conditions 1. and 2. above) whilst information from other wavebands
(conditions 3. and 4.) is used as supporting evidence. The origins of the radio emission from the seven 8.4-GHz-selected
sources with X-ray counterparts are less certain as they are
unresolved in the radio and have approximate spectral indices or upper
limits only. The \emph{ACS} images show that J123637+621135,
J123639+621249 and J123648+621427 are associated with spiral galaxies
with knots of star-formation (the lower-resolution CFHT image of J123648+621427
appeared elliptical).  J123644+621249 is associated with a pair of
apparently interacting optical galaxies at very similar redshifts. All
four have $\alpha\ga0.6$ and we list them as starbursts although the closest
(J123637+621135) is in fact of low luminosity, more like a normal
star-forming galaxy.  The remaining 3 have flatter spectra;
J123655+621311 is associated with an elliptical galaxy likely to
contain an AGN; the other two are unclassified.

In total, the 92 radio-bright sources include 23 unclassified objects,
52 starbursts and 17 AGN, using the radio-based classification. The 55
sources with X-ray counterparts include 9 of the unclassified sources,
36 starbursts and 12 radio AGN.  The starburst:AGN ratio is almost
identical, $\approx3:1$, to the that of the full radio-bright sample.
The three starbursts which contain radio AGN (counted once only, as
starbursts) are all X-ray detections. MIR observations only cover part
of the field but contain 22 sources detected at 15- or 16-$\mu$m as
well as in the radio and X-ray.  Nineteen of these (86\%) are radio
starbursts, including 3 with radio AGN cores, 4 with X-ray selected
type-II AGN (see Section~\ref{xorig}) and one with both. Two more are
probably AGN, J123646+621404 (also an X-ray type-II AGN) and
J123709+620841 (see \citealt{Muxlow05}). J123655+620808 is
unclassified as, although the \emph{ACS} image shows an apparently
spiral galaxy with a dust lane, the extended radio emission is
misaligned.

\subsection{Origins of X-ray emission}
\label{xorig}
The great majority of X-ray sources in the HDF region are unresolved
by \emph{Chandra} so only luminosity and spectral index information
may be available to determine the specific origin of the X-ray
emission.  Many classifications in the literature are based on optical
and other properties which could be due to separate mechanisms within
the host galaxy.  A comprehensive source-by-source breakdown is not
available but out of the 19 sources from the 1 Ms sample cross-matched
by \citet{Bauer02}, about 1/3 are classed as emission-line galaxies
and presumed to have X-ray emission of starburst origin; most of the
remainder are X-ray AGN.  

Star-forming galaxies and ULIRGs show a
close correlation between their star formation rates represented by
FIR emission, and both hard- and soft-band X-ray emission
\citep{Ranalli03}, although \citet{Rosa07} has recently found that, for
a higher-redshift sample in the CDF(S), the SFR implied hard-band
luminosities can be excessive compared from the rates derived from
soft-band or \emph{Spitzer} MIR data, presumably due to obscured-AGN
contamination in the hard band.  Hard X-ray emission associated with
star formation is thought to originate from high-mass X-ray binaries
\citep[e.g.][]{Grimm03}.  More slowly-evolving low-mass X-ray binaries
are likely to be less significant \citep{Rosa07}, especially in young
starburst galaxies. This leaves young supernova remnants and hot
plasmas associated with star-forming regions and galactic winds as
possible sources of the soft-x-ray component \citep{Ranalli03},
especially if super star clusters are forming \citep{Griffiths00}, as
discussed by \citet{Norman04}.

 The X-ray luminosity of most optically classified starbursts is
$<10^{35}$ W \citep{Alexander02XI} whilst the presence of detectable
hard band (2--8 keV) emission and X-ray luminosities $\ge10^{35}$ W is
usually taken to indicate the presence of an AGN; $\ge10^{37}$ W
implies a QSO \citep{Alexander03XIV}.  However, it is not unreasonable
that the most extreme starbursts could exceed an X-ray power of
$10^{35}$ W, if the X-ray luminosity is proportional to the rest-frame
IR emission \citep[e.g.][]{Ranalli03}, whilst some nearby FR~1 have
X-ray luminosities of only $10^{33} - 10^{35}$ W \citep{Evans06}.
Soft-band dominated X-ray emission (photon index $\Gamma\sim2$) can
indicate a starburst origin \citep{Ptak99} but is also seen from
unobscured AGN \citep{George00}. In the latter situation the emission
could be due to accretion or to jets but both mechanisms are
AGN-powered and included in X-ray AGN statistics.

Obscured (type II) AGN have a harder photon index ($\Gamma\la1.0$);
they are the only known sources with $\Gamma\la0.5$
\citep{Alexander05a} but $0.5<\Gamma<1$ is also seen from high-mass
X-ray binaries in starbursts.  Nonetheless, the combination of
$\Gamma\la1$ with a rest-frame 2--8-keV luminosity $L_{\mathrm{XH}}
\ge 10^{35}$ W can only be explained by a type-II AGN (see
Section~\ref{XH}).  \citet{Padovani04} identified a total of 91 such
sources in the HDF(N) with a hardness ratio corresponding
approximately to $\Gamma\la1.15$.  Of these, 64 lie within the
10-arcmin field and 17 are radio-bright. These are identified in
Table~\ref{tab2}.  A column density $N_{\mathrm{H}} > 10^{27}$ m$^{-2}$ is required to
provide sufficient obscuration. The estimates of $N_{\mathrm{H}}$
given by \citet{Alexander05a} for SMG confirmed that all 8 of the
type-II AGN common to their sample and ours exceed this threshold.

\citet{Bauer04} find that around 75--90\% of the 2-Ms HDF(N) X-ray
sources are AGN, of which about 2/3 appear absorbed, and about half
the remainder are starbursts.  A variety of studies of the GOODS
fields, including the use of multi-wavelength properties
\citep{Hornschemeier01, Bauer02, Szokoly04} give similar results,
implying a ratio of approximately 8:1 AGN to starbursts among the
X-ray detected sources.  

Our sample contains 62 X-ray sources with radio-bright counterparts of
which 17 or 18 appear to be heavily obscured X-ray AGN. In total, 37
(about 2/3) of the sources with measured redshifts, have hard-band
X-ray luminosities brighter than $10^{35}$ W (see Section~\ref{XH})
suggesting the presence of an AGN of some type \citep{Alexander03XIV,
Cowie04a}.  Statistically, the majority of all the X-ray emission is
probably AGN-powered but it is not possible to distinguish between
very luminous but softer emission from starbursts or unobscured AGN on
the basis of X-ray properties alone; moreover, diagnostics from other
regimes do not guarantee that the emission is from the same phenomenon
on a sub-galactic scale. We therefore concentrate on comparing the
X-ray-selected type-II AGN population with radio sources classified as
AGN or as starbursts.

\subsection{The high-redshift source J123642+121331}
\label{1331}

The highest redshift radio source, J123642+621331, at $z=4.424$, has a
high total 1.4-GHz flux density (467~$\mu$Jy). It has a steep radio
spectrum, it is a very reddened NICMOS detection \citep{Waddington99}
and it was detected by \emph{ISO} at 15~$\mu$m, all properties
consistent with starforming activity.  Its 1.4-GHz/FIR ratio, however,
is 20--50 times higher than other HDF(N) star-forming galaxies
\citep{Garrett02}.  The MERLIN+VLA image shows that about 10\% of the
flux is diffuse and extended at between 100--200 mas from the core
($\approx1$ kpc), which is likely to contain the starburst component.
The star formation rate inferred from the IR flux density is
$\approx1000$ M$_{\odot}$~yr$^{-1}$, comparable to the other highest
star formation rates deduced for starbursts in the HDF(N), which would
account for $\approx1$\% of the radio emission.

The compact core is AGN-dominated; the EVN detected over half the
1.4-GHz flux \citep{Garrett02} and global VLBI \citep{Chi06} resolves
a jet component a few tens of pc from the core.  J123642+621331 has a
measured $\Gamma=1.35$, above the limit for type-II AGN, but
\citet{Padovani04} noted that high-redshift sources might be
misclassified.  The expressions in
Section~\ref{sec:lx} assume that $\Gamma$ is constant from the
observed frame to the rest frame.  \citet{Alexander05a}, in their fig.~7,
demonstrate how absorption is a strong function of wavelength, such
that for $N_{\mathrm{H}} \ga 2\times 10^{27}$ m$^{-2}$, rest-frame
energies $\ga6$ keV are much less affected than lower energies.  At
$N_{\mathrm{H}} \ga 5\times10^{27}$ m$^{-2}$ the iron emission and
absorption lines, at rest frame energies 6--7 keV and 7--8 keV
respectively, become more prominent. The measured $\Gamma$ is derived
from the ratio of flux densities above and below 2 keV in the
observed frame.  This dividing energy corresponds to $\ge6$ keV at
$z\ge2$, so the observed $\Gamma$ of a high redshift absorbed source
may appear greater than the actual rest-frame 0.5--8 keV photon index.
In turn, the actual $L_{\mathrm{X}}$ would be higher than the value
given in Table~\ref{tab2}.  J123642+621331 would be most strongly
affected. If it is a type-II AGN with a rest frame 0.5--8 keV
$\Gamma\la1.15$, this is compatible with the observed $\Gamma=1.35$.

\section{Relationships between X-ray and radio luminosities}
\label{sec:relate}
\begin{figure*}
\sidecaption
\includegraphics[width=11cm]{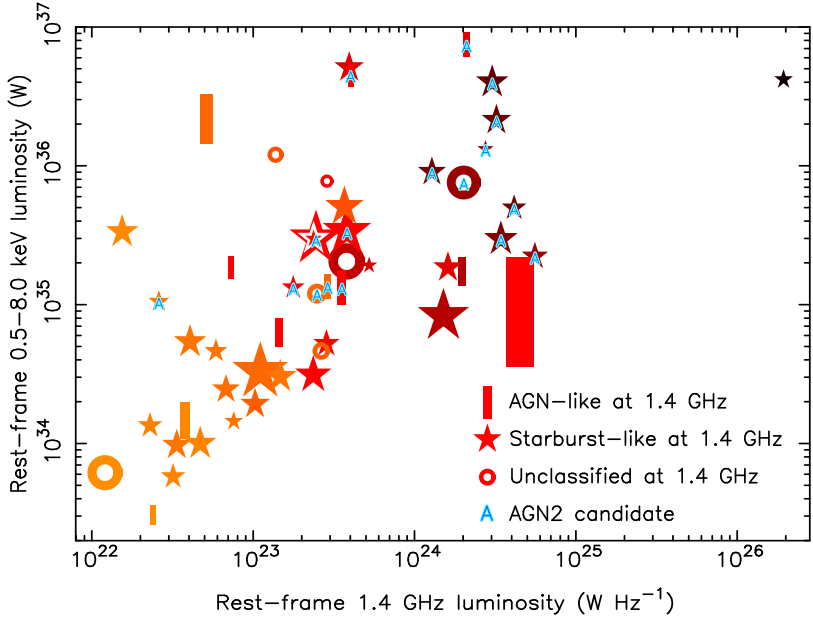}
   \caption{The distribution of the classes of radio-bright sources
           (see key) with respect to $L_{\mathrm{R}}$ and
           $L_{\mathrm{X}}$. The shade of red is proportional to
           redshift. The paler, brighter and darker symbols represent
           the approximate redshift ranges ($z<1$), ($1 < z < 2$) and
           ($z>2$), respectively. The size of the symbols is
           proportional to the source largest angular size.  The blue
           `{\bf A}'s represent X-ray selected type-II AGN; the
           highest redshift source may also be in this category.
           Three sources marked as starbursts also contain radio AGN
           (not shown to avoid overcomplicating this plot), see
           Fig.~\protect{\ref{LRLXBAUER.PS}}. Sources not detected at
           1.4 GHz are not shown as they have no radio angular size
           measurements.  }
      \label{LRLX_CZ.RPS}
\end{figure*}

\begin{figure*}
\sidecaption
\includegraphics[width=13.5cm]{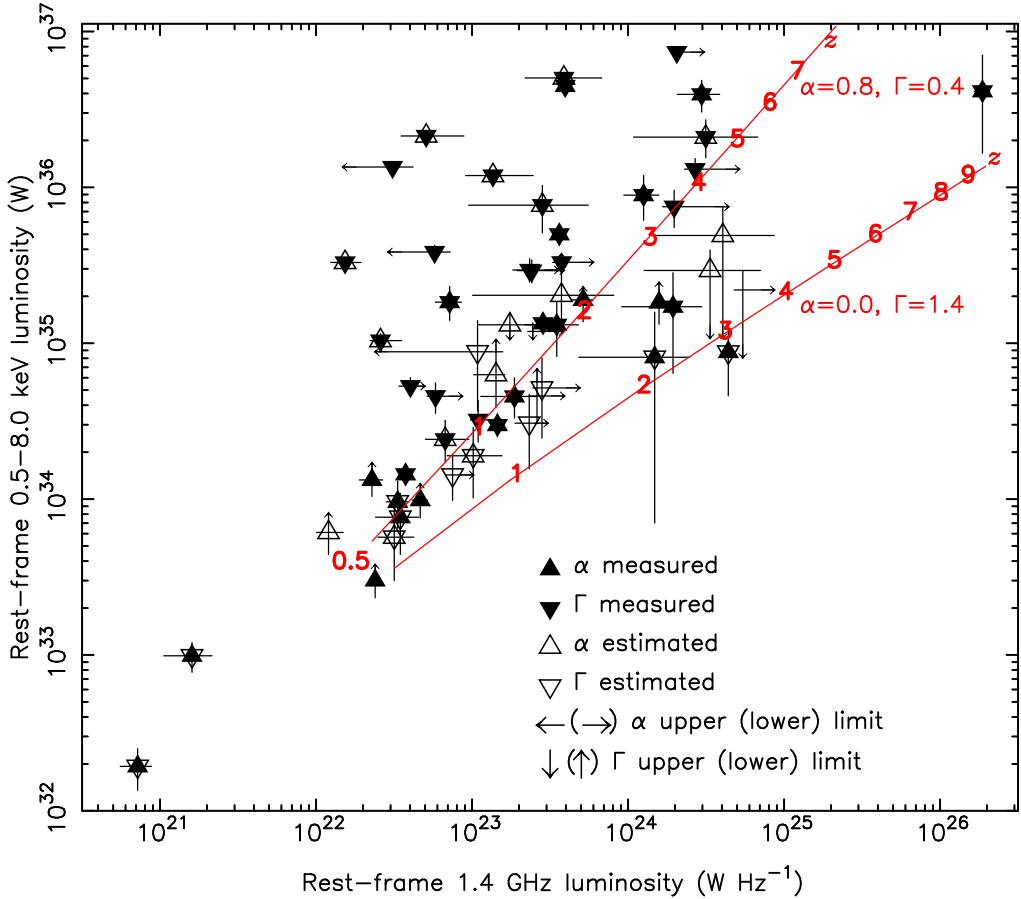}
  \caption{The accuracy of $L_{\mathrm{R}}$ and $L_{\mathrm{X}}$
           measurements. Uncertainties due to the spectral or photon
           index measurements are distinguished as shown in the key.
           See text for further explanations. The red lines show the
           detection limits at various redshifts for sources with the
           given combinations of $\alpha$ and $\Gamma$.  For example,
           a source with $L_{\mathrm{R}} = 10^{24}$ W m$^{-2}$ is
           detectable at $z<2$ if it has $\alpha\sim 0$, but would be
           seen out to $z\sim2.5$ if it had a steep radio
           spectrum. Its X-ray counterpart would be detected at
           $z\sim2$ if it had $L_{\mathrm{X}} > 5\times10^{34}$ and
           $\Gamma > 1.4$, but would need to have $L_{\mathrm{X}} >
           10^{35}$ if it had $\Gamma < 0.4$.  }
      \label{LRLXERR.RPS}
\end{figure*}

 Figure~\ref{LRLX_CZ.RPS} shows the relationship between $L_{\mathrm{
   R}}$ and $L_{\mathrm{X}}$ taken from Table~\ref{tab2} for the
   radio-bright X-ray sources with redshifts. The symbol sizes
   and shapes represent the largest angular size and the
   classification applied to the radio emission with a blue {\bf A}
   denoting the presence of an X-ray selected type-II AGN.  The shade
   of the symbols indicates the redshift.  The accuracy of our
   estimates of $L_{\mathrm{R}}$ and $L_{\mathrm{ X}}$ and potential
   selection effects are shown in Fig.~\ref{LRLXERR.RPS}, for all
   cross-matched sources with redshifts.  All sources shown have
   measured flux densities in both radio and X-ray regimes and
   measured redshifts (Section~\ref{x} and Table~\ref{tab1}).  Filled
   triangles pointing up (down) indicate objects which were detected
   at both 1.4 and 8.4 GHz (in at least two X-ray bands) giving a
   measured spectral (photon) index. Thus, the filled stars show
   objects with complete measurements.  Where one band is a defined
   limit, an arrow shows the resulting direction of uncertainty in the
   luminosity.  Open triangles show sources where the spectral or
   photon index has been estimated as described in Sections~\ref{1.4}
   and~\ref{x}.  The error bars were derived as described in
   Section~\ref{sec:lum} which also describes the method for
   estimating spectral (photon) indices where a source was only
   detected in one band in the radio (X-ray) regime; such sources are
   shown by open triangles pointing up (down).

Sources with a given rest-frame luminosity can
be detected at higher redshifts if they have steeper spectra, implying
that the sample might be biased towards radio starbursts with
low-obscuration X-ray counterparts. The red lines in Fig.~\ref{LRLXERR.RPS} show the limits of
detectability by the MERLIN+VLA and \emph{Chandra} observations
described in Section~\ref{data} for the two arbitrary combinations of
$\alpha$ and $\Gamma$. The lines are
marked with the redshifts out to which a source would be detectable
for the combination of luminosities at that point and the
spectral/photon index combination for that line.  This shows that
X-ray sources at $z\ga1.5$ need to
have $L_{\mathrm{X}}\ga10^{35}$ W (i.e. in the AGN regime) to be
detectable if they have harder photon indices.

\begin{figure*}
\sidecaption
\includegraphics[width=12cm]{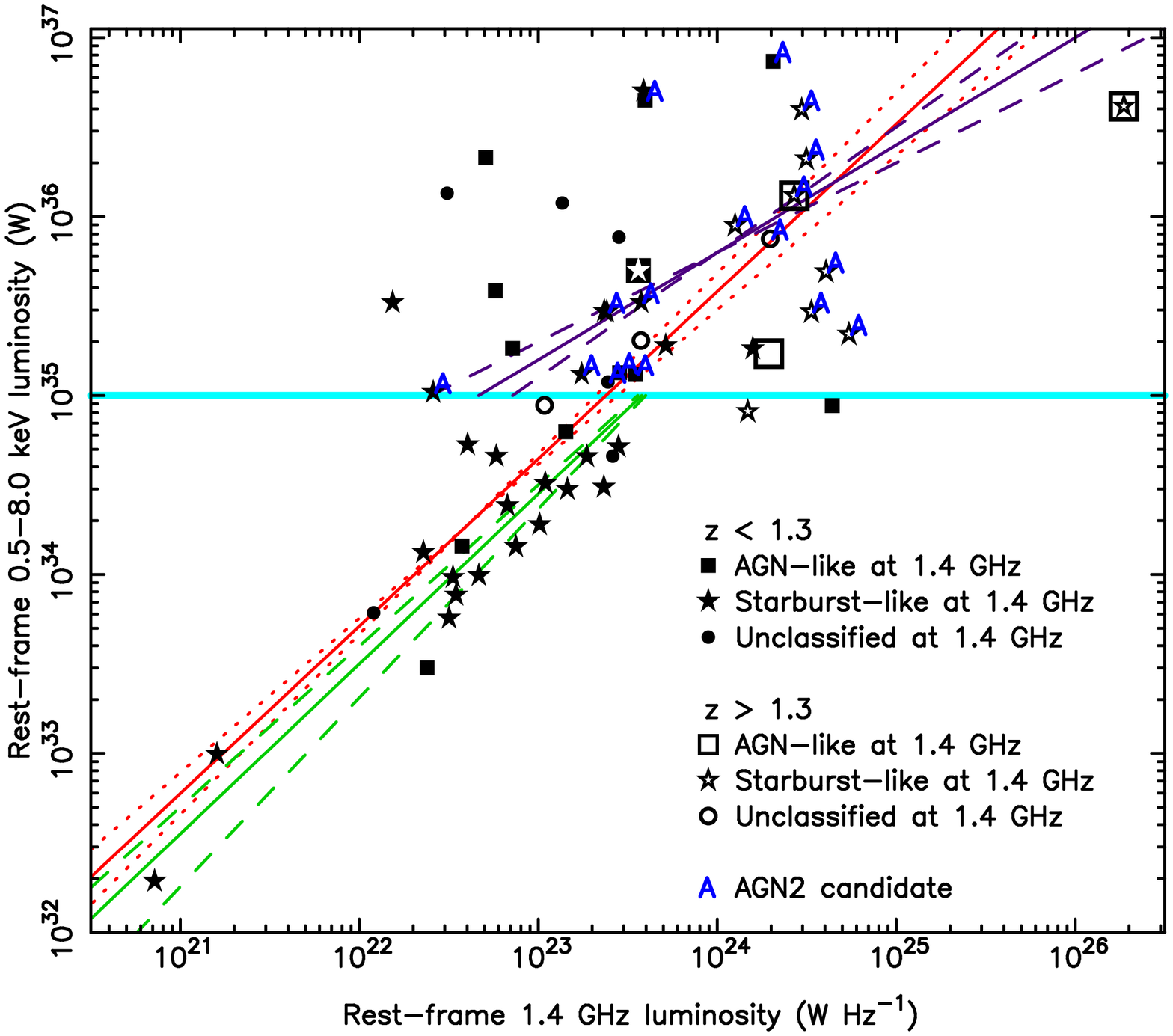}
   \caption{X-ray luminosity as a function of radio luminosity for all
           57 radio sources with X-ray counterparts and measured
           redshifts, classified as shown in the key (the squares
           enclosing stars are the three starbursts with apparent
           radio AGN cores; the lowest luminosity one is at
           $z<1.3$). The X-ray luminosity limit for AGN,
           $L_{\mathrm{X}}\approx10^{35}$ W, is marked by a horizontal blue dividing
           line.  The long red line is the relationship found by
           \protect{\citet{Bauer02}} (Equation~\protect{\ref{Bauer}}).
           The short green line shows the relationships for radio
           starbursts at $z<1.3$ with $L_{\mathrm{X}}<10^{35}$ W
           (Equation~\protect{\ref{loxloz}}). The short purple line
           shows the relationship for radio starbursts hosting X-ray
           selected type-II AGN (Equation~\protect{\ref{lxlragn2}}) at
           $L_{\mathrm{X}}>10^{35}$ W. The dashed lines show the
           uncertainties for each relationship. The blue {\sf A}s
           indicate X-ray selected type-II AGN; the most
           radio-luminous sources may also be in this category.  }
      \label{LRLXBAUER.PS}
\end{figure*}

\subsection{Relationships for starburst-selected sources}
\label{empire} 
Fig.~\ref{LRLXBAUER.PS} shows that there is a correlation  between
$L_{\mathrm{X}}$ and $L_{\mathrm{R}}$ at lower luminosities but there
is a very large scatter at $L_{\mathrm{X}}\ga10^{35}$ W, in particular
for sources at $z\ga1.3$. We investigated this by
looking for power-law relationships between $\log L_{\mathrm{ R}}$
and $\log L_{\mathrm{X}}$ expressed as
\begin{equation}
\log L_{{\mathrm{X}}} = B\log L_{{\mathrm{R}}} + A
\end{equation}
using values of the intercept and slope, $A_i$ and $B_i$ such that the
derived $i$th values of $\log L_{{\mathrm{X}}i}$ and $\log
L_{{\mathrm{R}}i}$ enclosed the observed values of $\log
L_{\mathrm{R}}$ and $\log L_{\mathrm{X}}$. We varied $A_i$ and $B_i$
between 1.0--50.0 and 0.1--5.0, respectively, in increments of 0.1. In
each case, we calculated the chi-squared:
$[(\log L_{{\mathrm{R}}i}-\log L_{\mathrm{R}})^2 + (\log L_{{\mathrm{X}}i}-\log L_{\mathrm{X}})^2]/\epsilon_i^2 $
where $\epsilon_i^2$ is the combined error taking into account asymmetric errors, e.g. using $L_{\mathrm{R-}}$ where
$L_{\mathrm{R}} > L_{{\mathrm{R}}i}$ and so on. 
We attempted to find values of
$A$ and $B$ corresponding to optimum values of the reduced
chi-squared, $\chi^2_i$,
for the whole data set and for various subsamples (e.g. sorted by radio or
X-ray class).

We obtained values of $\chi^2_i>5$ (very low significance) for
most data selections including all those based on radio-selected AGN.
The 10 radio-selected starbursts hosting X-ray-selected
type-II AGN gave the most significant result, with $\chi^2_i\sim3$
\begin{equation}
\label{lxlragn2}
\log L_{\mathrm{X}} = (0.6\pm0.1)L_{\mathrm{R}} + (21.4\mp2.4)
\end{equation}
(the uncertainties show the range in which the minimum $\chi^2$ was
obtained), shown by the purple line in
Figure~\ref{LRLXBAUER.PS}. Including J123642+621331 as a radio
starburst plus AGN (Section~\ref{1331}) gives an almost identical
relationship, consistent with the suggestion that it may also host a
type-II AGN, see Section~\ref{xorig}.  In any case, Tables~\ref{tab1}
and~\ref{tab2} show that the lower error bound to $\Gamma$ gives
J123642+621331 a flat enough photon index to be a type-II AGN and the
corresponding value of $L_{\mathrm{X}}$ still fits within the
uncertainties of Equation~\ref{lxlragn2}.

We obtained $3 < \chi^2_i < 5$ for low-luminosity, low-redshift
starbursts, e.g. $z < 1.3$, $L_{\mathrm{X}} < 10^{35}$ W (17 sources) gives 
\begin{equation}
\log L_{{\mathrm{X}}} = (0.95\pm0.05)\log L_{{\mathrm{R}}} +
12.6\mp1.2 
\label{loxloz}
\end{equation}
shown by the green line in Figure~\ref{LRLXBAUER.PS}.  These
coefficients are similar to those of Equation~\ref{Bauer}
\citep{Bauer02}, shown by the red line in Fig.~\ref{LRLXBAUER.PS}.
The slight systematic offset to lower X-ray luminosities in our sample
is probably because we used the 2 Ms data and a default of
$\Gamma=1.4$ where it was not measured. The low-redshift-dominated
sample of \citet{Bauer02} has a higher average measured photon index
and they took $\Gamma=2$ for sources which did not have a measured
value in the 1 Ms data. Six objects are common to both samples.

\citet{Barger07} dispute the existence of the radio-X-ray luminosity
  relationship, as applied to high-redshift samples including the
  HDF(N), suggesting that it is a selection effect.  We cannot compare
  this directly with our weak correlation given in
  Equations~\ref{lxlragn2} and \ref{loxloz} since about half of our
  radio starbursts with obscured X-ray AGN and a third of low-X-ray
  luminosity starbursts have $S_{\mathrm{R}}$ less than their cutoff of
  60 $\mu$Jy, and our criteria for starburst classification is more
  specific to the origins of the radio emission than is their optical
  method.  The significant point for both this paper and
  \citet{Barger07} is that although a high proportion of high-redshift
  star-forming sources detected in the radio are also detected in
  X-rays, their luminosities are weakly correlated or uncorrelated,
  suggesting that the X-ray emission is of non-starburst origin.

Figure~\ref{LRLXBAUER.PS} also shows that 2 out of 4 radio AGN with
$L_{\mathrm{X}} < 10^{35}$ W lie close to the starburst-based
relationships and the other 2 are under-luminous in X-rays (although
one of these, the source with the lowest value of
$L_{\mathrm{X}}/L_{\mathrm{R}}$, is the FR~1 J123644+621133, the only
radio-X-ray crossmatched source to have jet-dominated radio emission).
Conversely, only one each of a radio starburst and an unclassified
source are present with $z>1.3$ and $L_{\mathrm{X}} < 10^{35}$ W
although as shown by Fig.~\ref{LRLXERR.RPS}, such a source detected in
one regime would also be detected in the other out to $z\sim2$ for a
typical starburst spectral index.

We emphasise that the significance of quantitative luminosity
relationships is low for samples drawn exclusively from the HDF(N)
data.  The presence of X-ray and radio emission appears to be
correlated (Section~\ref{z}), but not the precise luminosities,
whether starbursts or any other classes of objects are studied.  The
clear correlation between $L_{\mathrm{ X}}$ and $L_{\mathrm{R}}$ found
for samples dominated by lower-redshift sources does not apply to the
high-redshift, high-luminosity sources.  We already noted (in
Section~\ref{z}) that the nature of the overlap between radio and
X-ray detections changes around $z\sim1.1$, which supports our
contention that 
the relationship between radio- and X-ray-emission mechanisms changes
dramatically around $z=1.1-1.3$.

\subsection{Hard-band X-ray luminosity}
\label{XH}

\begin{figure*}
\sidecaption
\includegraphics[angle=0,width=12cm]{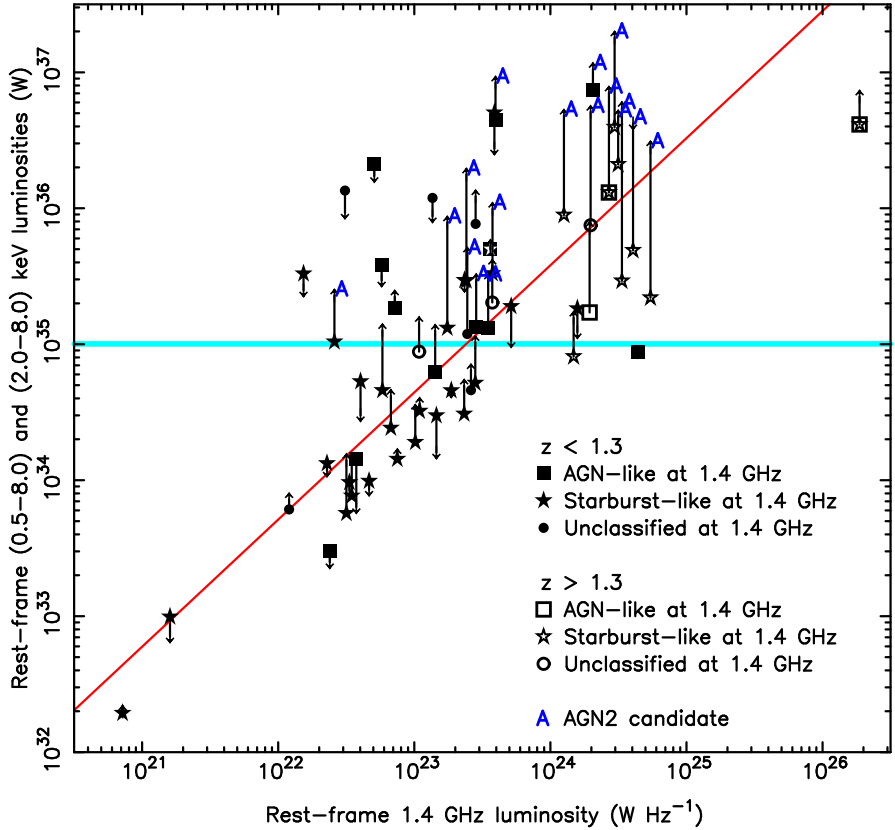}
   \caption{The symbols, dividing line at $L_{\mathrm{X}}\approx10^{35}$ W and red line are as in
     Fig.~\protect{\ref{LRLXBAUER.PS}}. The arrow tips show the X-ray
     luminosity recalculated for the 2--8 keV rest-frame,
     $L_{\mathrm{XH}}$, derived from the hard band flux densities
     using a constant $\Gamma=1.8$ in
     Equation~{\protect{\ref{LumX}}}. The X-ray selected type-2 AGN are
     labelled at the $L_{\mathrm{XH}}$ values only.}
      \label{LRLXHB.PS}
\end{figure*}

The values of $L_{\mathrm{X}}$ given in Table~\ref{tab2}, derived
 using Equation~\ref{LumX} in Section~\ref{sec:lx}, do not take into
 account corrections for absorption nor for the intrinsic photon
 index.  \citet{Padovani04} derived the rest-frame X-ray luminosities
 for candidate type-2 AGN using the observed hard-band flux densities
 and assuming an intrinsic $\Gamma=1.8$. We follow this method to
 calculate $L_{\mathrm{XH}}$ for all the cross-matched sources with
 published redshifts, shown as the arrow ends in Fig.~\ref{LRLXHB.PS}.
 We looked for relationships between $L_{\mathrm{XH}}$ and
 $L_{\mathrm{R}}$ using a method similar to that described in
 Section~\ref{empire} but btained results of even lower significance.

Figure~\ref{LRLXHB.PS} shows that all the type-2 AGN have
$L_{\mathrm{XH}}>L_{\mathrm{X}}$, as expected, and that the difference
is greater for the more radio-luminous sources.  All sources with
arrow tips above the blue horizontal line have $L_{\mathrm{XH}}>10^{35}$ W,
indicating the presence of an X-ray AGN \citep{Alexander03XIV, Cowie04a}.

We used Equation~\ref{Bauer} to predict the full-band X-ray luminosity
of starburst origin, $L_{\mathrm{XSB}}$, assuming that
$L_{\mathrm{R}}$ was entirely due to starburst activity.
$L_{\mathrm{XSB}}$ is an overestimate where a significant fraction of
radio emission is of AGN origin, although J123642+621331 is the only
powerful starburst where over half the radio emission is AGN-powered
(Section~\ref{radorig}). As it has been shown empirically
\citep[e.g.][]{Ranalli03} that hard- and soft-band X-ray emission of
starburst origin have similar dependencies in the relationships with
emission of common origin in other regimes, the difference between the
measured hard-band luminosity and $L_{\mathrm{XSB}}$ can thus be
regarded as a lower limit to the X-ray luminosity of non-starburst origin.  We
find that $(L_{\mathrm{XH}} - L_{\mathrm{XSB}})$ exceeds $10^{35}$ W
for all 17 type-2 AGN identified by \citet{Padovani04}, including the
10 radio starbursts, with a mean difference of $\ga3.9\times10^{36}$ W
(much greater than the total uncertainties) for either sub-sample.  We
interpret this as confirming that these galaxies must contain AGN,
since the excess X-ray power, $\gg10^{35}$ W, is very unlikely to be be
explained by any other mechanism.  These values also suggest that, on
average, $\la1/3 $ of X-ray emission from obscured type-2 AGN is of
starburst origin.  All the radio starbursts with type-2 AGN remain
sufficiently X-ray-bright to meet the AGN selection criteria, even
using the lower luminosity limit derived here.

 $L_{\mathrm{XH}} - L_{\mathrm{XSB}}$ is lower, at
$\approx1.6\times10^{36}$ W and $\approx2.8\times10^{34}$ W for all
radio-selected AGN and all starbursts respectively.  Starbursts
without type-II AGN show a very large scatter in the difference, about
a mean of $\approx-2.1\times10^{36}$ W, showing that any AGN
contribution is negligible in the majority of `pure' starbursts.

\section{X-ray-selected type-II AGN associated with radio-selected starbursts}
\label{sbagn}
\subsection{A distinct population}
\label{distinct}

 Radio starbursts hosting type-2 AGN dominate the radio
detections at high redshift, illustrated in
Fig.~\ref{ZX_AGN2_XRAGN2.eps}. 
The 10-arcmin field contains 64 X-ray selected type-II AGN (as defined
in Section~\ref{xorig}), plus the candidate type-II AGN J123642+621331
(Sections~\ref{1331} and~\ref{empire}). Eighteen of these have
radio-bright counterparts (none are among the additional 7 8.4-GHz
selected sources).  18/64 is a similar proportion to the quarter of
all X-ray sources in the 10-arcmin field which are type-II AGN
candidates.  Nine out of the 11 radio-bright X-ray
sources at $z>1.3$ host type-II AGN and this includes 8 out of the 9
radio starbursts in this redshift range.

\begin{figure}
\centering
\includegraphics[angle=0,width=8.8cm]{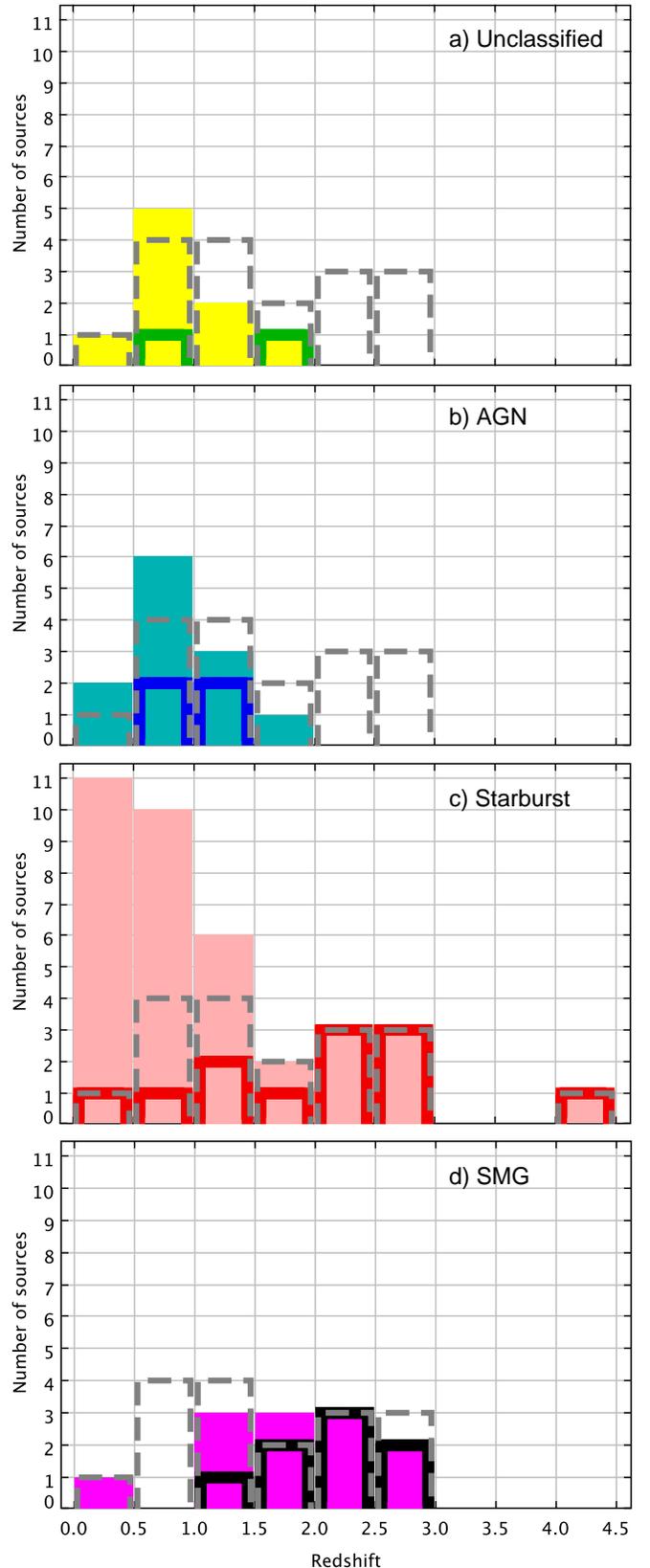}
   \caption{Redshift distribution of radio/X-ray sources according to
     radio classification and counterparts.  In each panel the dashed
     grey line shows the distribution of sources which have X-ray type-II AGN
     signatures (including J123642+121331,
     Section~\protect{\ref{1331}}) and the solid line shows
     their counterparts in the sub-sample.  The filled area shows the total number of sources
     with radio properties which are a) unclassified, b) AGN-like or
     c) starburst-like; panel d) shows radio sources with SCUBA
     counterparts.
}
      \label{UAGNSB_all_AGN2.eps}
\end{figure}

\begin{figure}
\centering
\includegraphics[angle=0,width=8.8cm]{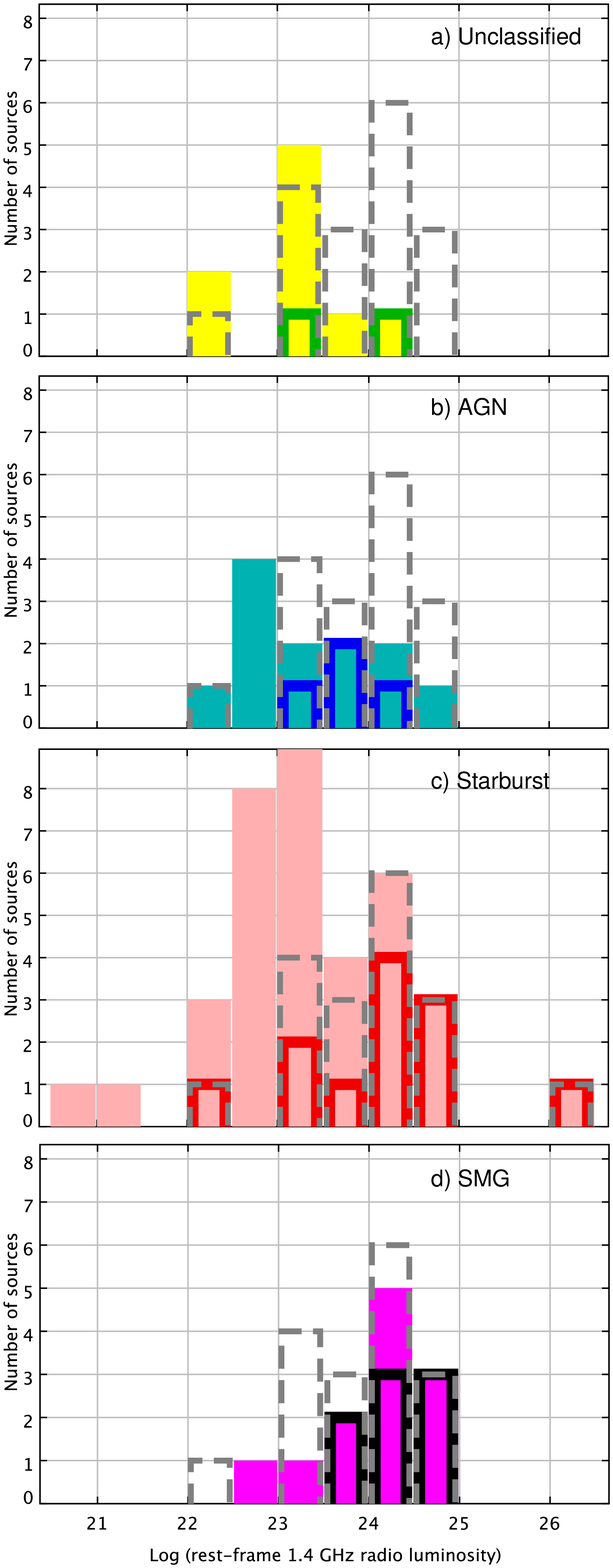}
   \caption{Radio luminosity distribution of radio/X-ray sources
     according to radio classification and counterparts, see
     Fig.~\protect{\ref{UAGNSB_all_AGN2.eps}} for more details.}
      \label{UAGNSB_all_AGN2_Lr.eps}
\end{figure}

Figure~\ref{LRLX_CZ.RPS} and Table~\ref{tab1} show that the majority
of the X-ray selected type-II AGN with radio counterparts are associated with
starbursts (classified using the criteria in Section~\ref{radorig}).
The breakdown by radio source type, redshift and luminosity is shown
in more detail in Figs.~\ref{UAGNSB_all_AGN2.eps}
and~\ref{UAGNSB_all_AGN2_Lr.eps}.  The 3 radio starbursts
(J123635+621424, J123642+621331 and J123642+621545) which also show
signs of containing radio AGN (Section~\ref{radorig}) are only shown
once, as starbursts.  J123635+621424, and probably J123642+621331, are
also X-ray selected type-II AGN.  

Radio AGN outnumber starbursts 2:1 in the most powerful third
($S_{\mathrm{R}}>100$ $\mu$Jy) of all 92 radio-bright sources whilst
starbursts are an even greater majority among the fainter sources.
The radio-selected AGN are on average intrinsically more
radio-luminous than the starbursts.  The median redshift of all radio
AGN is higher compared with all starbursts in the HDF(N) and they
appear to be separate populations \citep{Muxlow05}.  The high redshift
radio-selected starbursts associated with obscured X-ray selected AGN
appear to be a third class, distinct from the radio-bright AGN and
from the majority of the starbursts. 

Figure~\ref{GZ.PS} shows a
remarkable segregation between high-$\Gamma$, low-$z$, sources of all
classes but without type-II AGN, and low-$\Gamma$ sources.  Of the
latter, the radio starbursts and unclassified sources with type-II AGN
(including the candidate J123642+621331) show a slight correlation
between $\Gamma$ and $z$, consistent with the change in X-ray spectral
slope expected for highly obscured sources at higher rest-frame
energies (Section~\ref{xorig}). The blue $\Gamma(z)$ line is a very
rough estimate of the change in the observed $\Gamma$ with increasing
redshift, for a rest-frame $\Gamma=0$, based on the spectral models
shown in \citet{Alexander05a}, their fig.~7, for sources with
$N_{\mathrm{H}}\ga2\times10^{27}$ m$^{-2}$.  All but 3 out of 16
sources with a $\Gamma$ below the estimated $\Gamma(z)$ line contain
type-II AGN and the majority are radio starbursts; the only
radio-bright AGN here is not a type-II AGN.

The 4 radio-bright AGN with X-ray type-II AGN all lie above this line,
suggesting that they do not reach the highest degrees of obscuration.
The segregation is not due to Compton-thick absorption as signs of
this are only seen in one source, J123622+621629, a radio starburst
hosting a type-II AGN, \citep{Alexander03XIII}.


 Although all but 2 of the radio AGN with measured redshifts are
detected by \emph{Chandra}, the non-detections are the
highest-redshift sources; in contrast, all the starbursts known to
have $z>1.3$ have X-ray counterparts.  Figures~\ref{UAGNSB_all_AGN2.eps}
and~\ref{UAGNSB_all_AGN2_Lr.eps} (panels {\bf a} and {\bf b}) show
that the distributions of X-ray selected type-II AGN are biased towards much
higher redshifts and luminosities than the distributions of
radio-classified AGN or unclassified radio sources.  The population of
type-II AGN does overlap closely the high redshift/high luminosity end of the
distribution of radio-selected starbursts (panel {\bf c}).

\subsection{Radio-bright SMG starbursts with type-II AGN}
\label{scubasec}

There are strong similarities between the redshift distributions of
the radio SMGs and type-II AGN (Figs.~\ref{UAGNSB_all_AGN2.eps}
and~\ref{UAGNSB_all_AGN2_Lr.eps}, panel {\bf d}), especially at
higher $z$. 20\% of all radio-bright galaxies detected by
\emph{Chandra} are also SCUBA
sources (Section~\ref{scuba}) and nearly half of
these contain type-II AGN. 
We noted a
low correlation between $L_{\mathrm{ X}}$ and $L_{\mathrm{R}}$ for
high-redshift starbursts (Section~\ref{empire}); a similar large scatter was
seen by \citet{Borys04} for a SCUBA-selected subsample of 10 sources.

\citet{Alexander05a} calculated the radio and X-ray luminosities for
all SMG with spectroscopic redshifts, in the overlap between the
\emph{Chandra} and VLA fields of view. All except one have X-ray luminosities in excess of the
relationship predicted from star formation, by over an order of
magnitude in the case of the  type-II AGN candidates 
(as we also found in Section~\ref{XH}).
The X-ray selected
type-II AGN also have an 
 FIR/X-ray luminosity ratio  about an order of magnitude greater than the
typical ratio for nearby QSO, showing that at least 90\% of the FIR
emission is probably of starburst origin, at an SFR of order 1000
M$_{\odot}$ yr$^{-1}$.

\subsection{Extended radio emission, compact X-ray cores}

The starburst interpretation of sub-mm emission from SMG which are
hard X-ray sources is only questionable if they possess nuclear dust to
much greater optical depths than is seen around local AGN
(Section~\ref{intro}; \citealt{Alexander05a}).  We summarise the
evidence that these objects do contain extended
starbursts which are distinct from any AGN cores.

The radio starburst classification is based on distinctive morphology
and spectral index (Section~\ref{radorig}; \citealt{Muxlow05}).  The
median angular size of radio-bright AGN with X-ray counterparts is
$\sim$0\farcs6 whilst for starbursts it is $\sim$1\farcs4
(Table~\ref{tab1}; Fig.~\ref{LRLX_CZ.RPS}). SCUBA galaxies which are
most extended in the radio are also more likely to be X-ray bright; of
the 12 SMG radio sources studied by \citet{Chapman04a}, the 6 with the
largest 1.4-GHz angular sizes all had X-ray counterparts but only 2
of the 4 smallest radio sources had X-ray counterparts.

X-ray emission from many of the the starburst galaxies would,
however, be resolved, if it had a common origin with the radio
emission, which is not seen -- all 55 X-ray sources are smaller
than the 1\arcsec\/ \emph{Chandra} FWHM in this region.

Nine out of the 10  radio-bright X-ray sources which are
also SMG are radio starbursts with $\alpha\ge0.7$ (including 7 with
type-II AGN) and the tenth is unclassified.  The
mean radio angular size is 1\farcs3 and only J123635+621424, a
starburst with a radio AGN core, is smaller than 0\farcs6.
\citet{Chapman04a} found a similar result, obtaining a typical extent
of 1\arcsec\/ for the radio counterparts to a sample of dozen SMGs.
\citet{Pope05} found that the the optical \emph{ACS} counterparts of
SMG at $z<2$ in the HDF(N) have radii in the range $\sim$1\arcsec\/ --
$\sim$2\farcs5, which is significantly larger than field galaxies at
the same redshift.
Similarly, more distant ($z\approx2.5$), highly reddened galaxies with SEDs
consistent with vigorous star formation are typically over twice the
size of non-starformers in \emph{HST NICMOS} images of the HDF(S)
\citep{Zirm06}. 

We conclude that, although present instruments cannot resolve MIR or
 sub-mm sources on the scale of radio or optical emission with
 starburst characteristics, the close association supports a common
 origin.  In contrast, X-ray emission from the same sources is
 consistently more compact than the radio emission and appears to have
 a distinctly different origin.

\begin{figure}
\centering
\includegraphics[angle=-90,width=8.8cm]{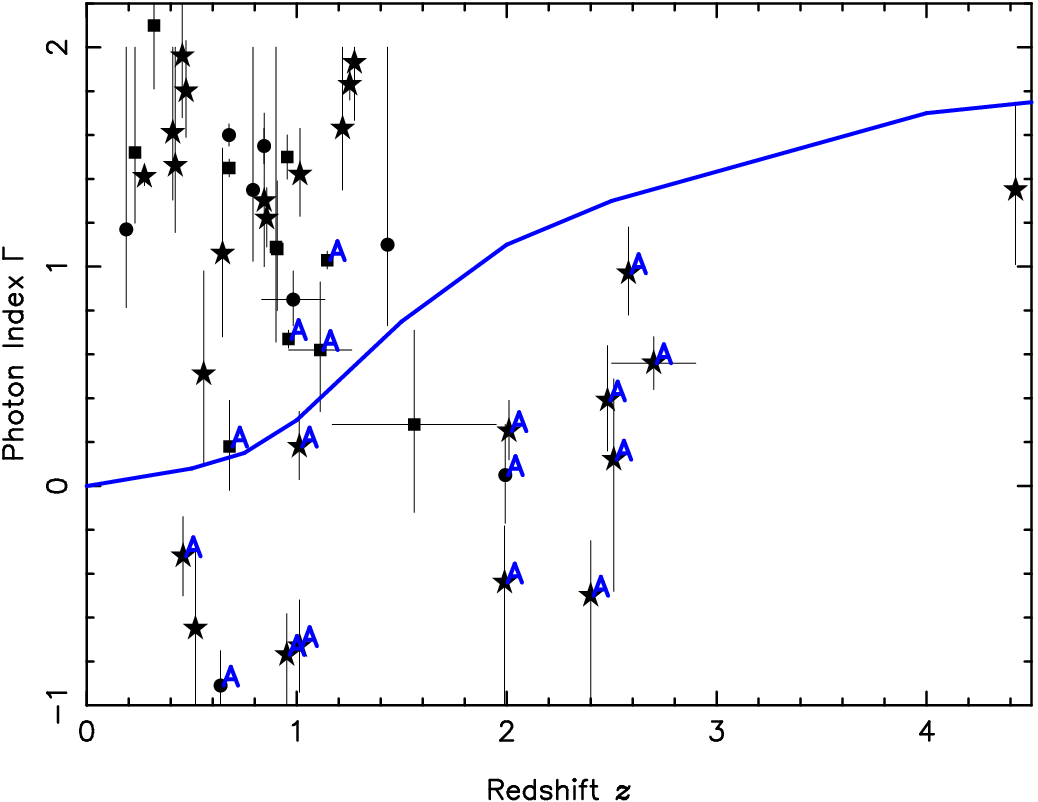}
   \caption{X-ray photon index as a function of redshift for radio
           starbursts (stars), AGN (squares) and unclassified sources
           (circles). All photon indices are fully measured or are
           partly measured upper or lower limits (see
           Section~\protect{\ref{sec:lx}}). X-ray selected type-II AGN are labelled
           {\sf A}. The blue line shows a rough
           extrapolation of an observed-frame photon index
           $\Gamma=0$ at $z=0$ out to $z=4.5$, using  estimates
           based on the models shown in \protect{\citet{Alexander05a}}
           for sources with $N_{\mathrm{H}} {\ge} 2\times 10^{27}$ m$^{-2}$.  }
      \label{GZ.PS}
\end{figure}

\section{The nature of very faint radio and X-ray sources}
\label{radio-faint}

At high redshifts, the present generation of radio and X-ray surveys
are biased towards extreme objects, as whatever was `normal' at
$z\gg1$ is below the threshold for secure detections of individual
sources.  In advance of \emph{e}-MERLIN, the EVLA and XEUS, stacking
is the only method to investigate the radio and X-ray properties of
objects at $z > 1$ with $L_{\mathrm{R}} < 10^{23}$ W Hz$^{-1}$ or
$L_{\mathrm{X}} < 10^{34}$ W.  \citet{Muxlow05} demonstrated the
existence of such a radio population by finding a statistical excess
of 1--6 $\sigma$ radio flux on arcsec scales at the positions of {\em
ISO} sources and of optical (\emph{HST WFPC2}) galaxies with \emph{I}
$\la 24$ mag in the 3-arcmin field. The MERLIN+VLA images are more
than twice as sensitive as the VLA-only data but no new sources were
detected in the 3-arcmin field between 27--40~$\mu$Jy, suggesting that
any objects just below the VLA completeness limit are extended over
$\ga2$\arcsec\/ with no hotspots.  \citet{Garrett00} showed that about
10\% of radio sources in the HDF(N) are probably resolved out even at
2\arcsec\/ resolution so even stacking MERLIN+VLA data will not
recover these.

We analyse the excess radio flux at the positions of \emph{Chandra}
sources in the recently-made 8-arcmin radio image convolved with a
restoring beam of 0\farcs4 (see Section~\ref{parsel}).  \citet{Muxlow06}
and \citet{Beswick06} present similar analyses at the
positions of \emph{ACS} and \emph{Spitzer} sources.  

\subsection{Radio-faint Chandra sources}
\label{radiofaint}
181 X-ray sources from the catalogue of \citet{Alexander03XIII} lie within
the 8-arcmin MERLIN+VLA 1.4-GHz image.  We used
the matching criteria described in Section~\ref{x} to divide the
X-ray sources into 39 sources with radio counterparts brighter than
40~$\mu$Jy (``radio-bright''), and 142 without (``radio-faint'').  For
each sample, we stacked the 1.4-GHz radio flux density enclosed by 0\farcs25 to
2\arcsec\/ radius circles centred on the X-ray source positions. We
constructed control samples by stacking the radio flux at positions
offset by 10\arcsec\/ from each X-ray source (checking that these did
not coincide with known X-ray or radio  sources or their sidelobes).  
The noise distribution  is slightly
non-Gaussian due to confusing sources in the sidelobes of the
heterogeneous primary beams but there are no
artifacts $>7\sigma$ ($\approx27$ $\mu$Jy per 0\farcs2 -- 0\farcs5
beam) apart from sidelobes very close to the 3 sources brighter than 1
mJy, J123644+621133, J123714+620823 and J123725+621128
\citep{Muxlow05}.  
The average stacked flux density from
the radio-faint sources increased up for radii up to $\sim1$\arcsec\/
and levelled off for higher radii, so we report results for 1\arcsec\/
radii. The variance in each bin is 3--4~$\mu$Jy
(the radio map $1\sigma$ level) and the control samples for the
radio-faint sources are all
$<3\sigma$.

\begin{figure}
\centering
\includegraphics[angle=-90,width=8.8cm]{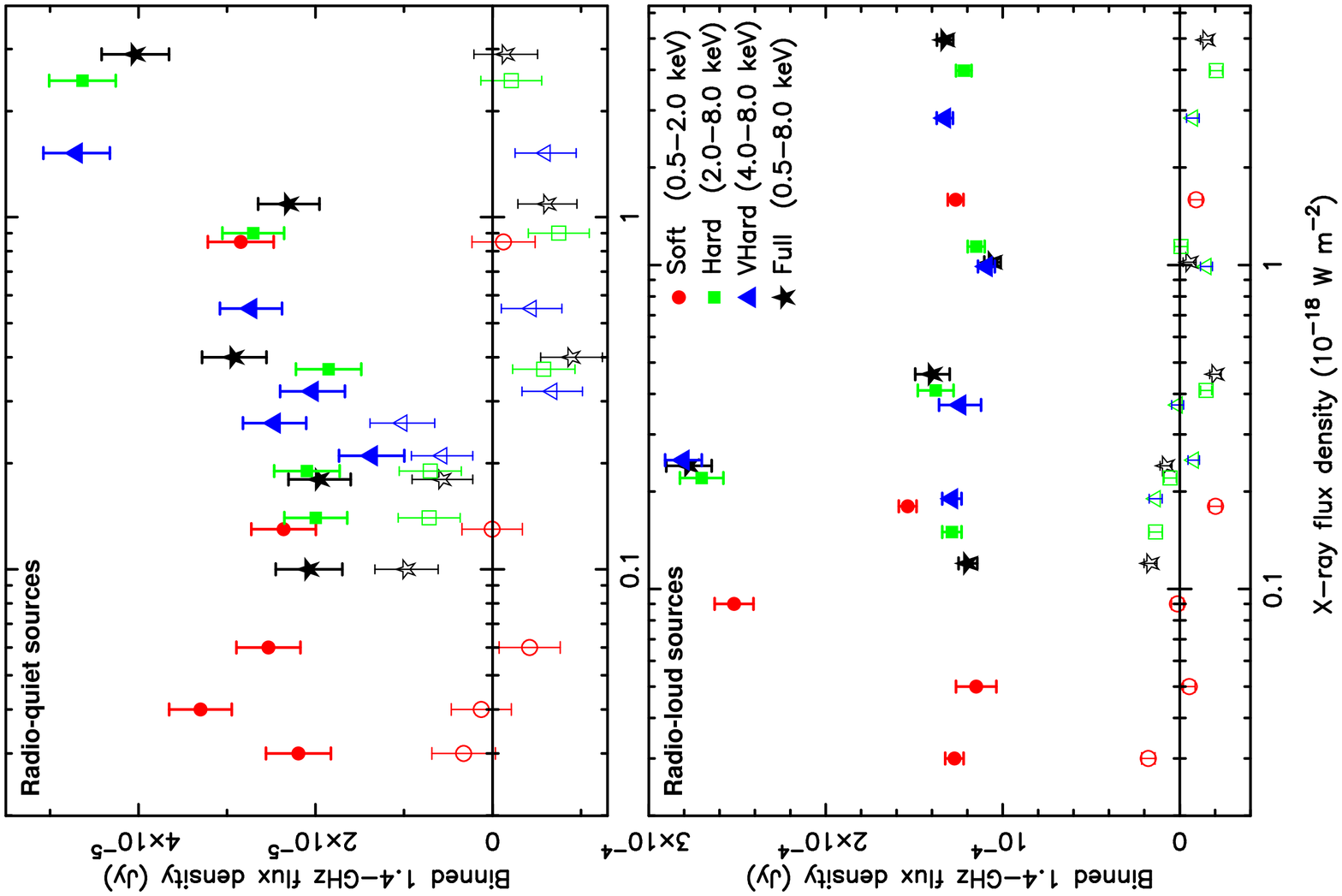}
   \caption{The heavy symbols in the upper and lower panes show the
           binned 1.4-GHz flux densities at the positions of radio-faint and
           -bright X-ray sources, respectively. This was measured
           separately for 4 X-ray bands, see key.  The ordinate
           axis shows the median X-ray flux in each bin for each
           band. The faint hollow symbols show the binned 1.4-GHz flux
           densities for
           control samples at 10\arcsec\/ separation from each X-ray
           source.  The error bars show the rms scatter in each bin.
           }
      \label{BINBYCXOFLUX.PS}
\end{figure}
\begin{figure}
\centering
\includegraphics[angle=-90,width=8.5cm]{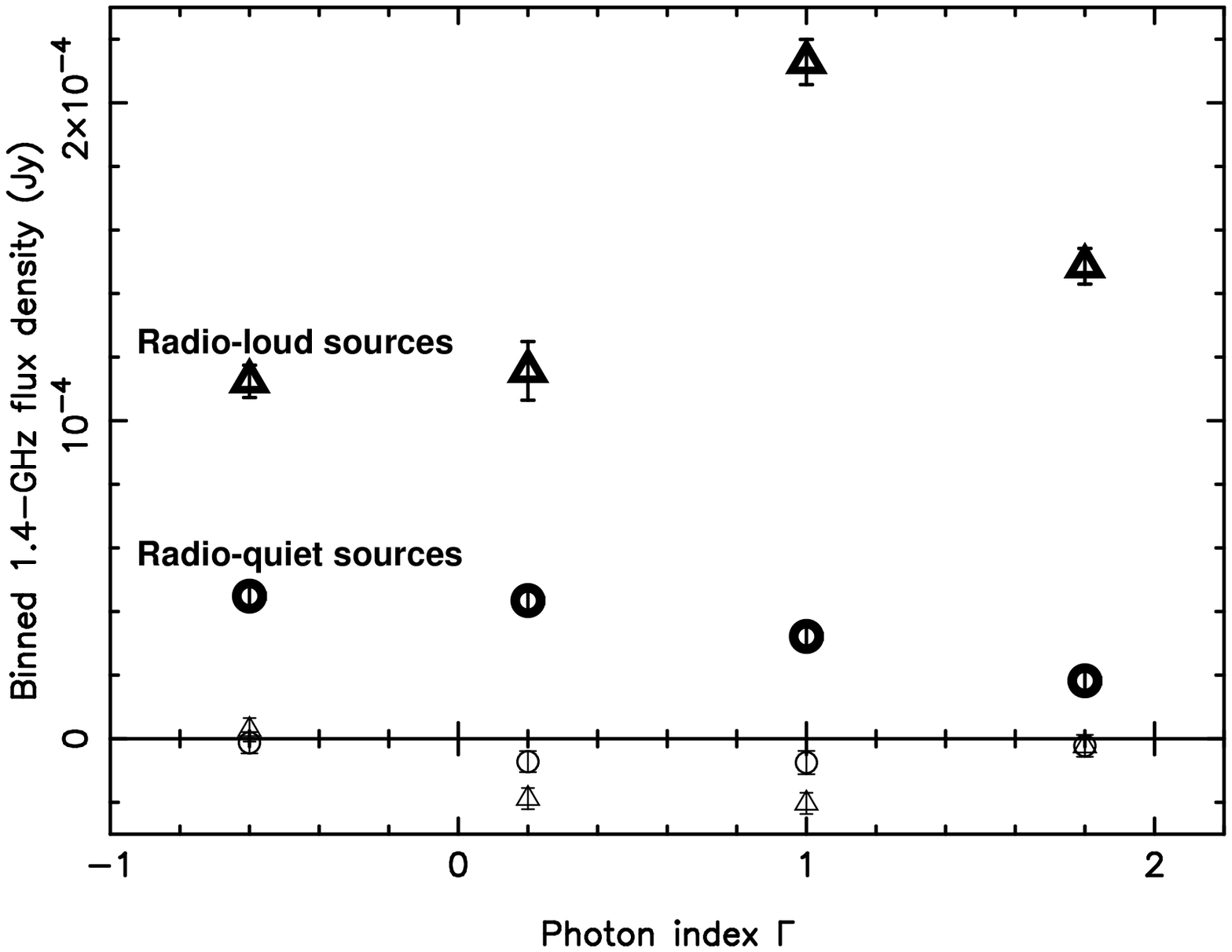}
   \caption{The heavy circles and triangles show the
           binned 1.4-GHz flux densities at the positions of radio-faint and
           -bright X-ray sources, respectively. The ordinate
           axis shows the median X-ray photon index in each bin for each
           band. The faint hollow symbols show the binned 1.4-GHz flux
           densities for
           control samples at 10\arcsec\/ separation from each X-ray
           source.  The error bars show the rms scatter in each bin.
           }
      \label{BINBYCXOGAMMA.PS}
\end{figure}
The stacking intervals were sorted by X-ray flux
density, photon index and redshift, into bins containing almost equal
numbers of measurements.  In each of
Figs~\ref{BINBYCXOFLUX.PS} and~\ref{BINBYCXOGAMMA.PS} the scale on the
abscissa shows the median value in each bin.

\subsubsection{Stacking by X-ray flux density}
\label{sec:stackf}
We present results for stacking sorted by full, soft, hard and very
hard (4--8 keV) band flux density.  We plot the relationship between the mean stacked radio
flux density and the 
 X-ray flux density in each of 5
bins with 28 or 29 (7 or 8) samples in each bin for the radio-faint
(-bright) sources. 
 We treated upper limits as values; hence the
nominal value for the bottom bin may be greater than the true median
X-ray flux but it does not affect the significance of the results.
The filled symbols in Fig.~\ref{BINBYCXOFLUX.PS}, upper panel, show the
stacked radio flux at the X-ray positions of radio-faint sources (see
key) and the hollow symbols show the corresponding control values.
This shows that the mean stacked radio flux density is significantly
above the control for all bins apart from the faintest very hard band
bin.  The average radio flux density is similar for any soft-band X-ray flux
density, but becomes increasingly correlated with X-ray flux density
for the hard and very hard bands. Figure~\ref{BINBYCXOFLUX.PS}, lower
panel, shows that there is no correlation with any band for
radio-bright sources (the four high values are biased by one very
bright radio source), as discussed in Section~\ref{sec:relate}.

\subsubsection{Stacking by photon index}
84 (29) radio-faint (-bright) X-ray sources have at least partly
measured photon indices ($\Gamma$) (i.e. detected in at least one
sub-band as well as full band). We plot the mean stacked radio flux
density against the X-ray photon index in each of 4 bins
evenly spaced in $\Gamma$, which gives similar source counts in each
bin.  Figure~\ref{BINBYCXOGAMMA.PS} shows that radio-faint X-ray
sources (heavy circles) with $\Gamma < 0.6$ have a mean radio flux
density twice that for
sources with $\Gamma > 1.4$.  This is not the case for the radio-bright
sources (heavy triangles).

\subsubsection{Stacking by redshift}
76 (35) radio-faint (-bright) X-ray sources 
have measured redshifts.  In each case we divided them into 4 bins
with approximately equal numbers of samples in each bin.  There is no
obvious correlation for either sample. 

\subsection{The nature of radio-faint X-ray sources}
\label{rfn}

There is a significant excess of radio flux at the positions of {\em
Chandra} sources without formally identified radio counterparts.
These have typical 1.4-GHz flux densities from $\approx20-40$~$\mu$Jy
per 1\arcsec-radius circle ($\equiv6-13$~$\mu$Jy arcsec$^{-2}$).  
This is the median largest angular size
of radio sources above $40$~$\mu$Jy \citep{Muxlow05} and is three
times the median \emph{Chandra} position error for the sources within
the 8-arcmin field (Section~\ref{x}).

The most obvious correlations for radio-quiet sources are the
 association of a higher average radio flux density with a higher
 X-ray flux density in the full or harder bands, and with a lower
 $\Gamma$.  The average radio flux density approaches 40~$\mu$Jy for
 radio-quiet X-ray sources with flux densities $>10^{-18}$ W m$^{-2}$
 Hz$^{-1}$ or with hard photon indices, suggesting that the majority
 of these (probably obscured) X-ray sources have genuine radio
 counterparts.  We examined the radio-quiet images at the position of
 each \emph{Chandra} source by plotting contours at 3, 4, 5 and 6
 $\sigma$\footnote{An image cut-out service will be available via AstroGrid.} Eleven sources are $>20$
 $\mu$Jy (5$\sigma$). Their apparently extended radio flux and the
 absence of hotspots brighter than 27~$\mu$Jy per 0\farcs4 beam,
 corresponding to maximum brightness temperatures of less than a few
 100 K, suggests that these sources are part of the radio starburst
 population and do not contain radio-bright AGN.  Seven of the
 corresponding X-ray objects have a measured photon index of which 5
 have $\Gamma<1$ and 4 are listed as type-II AGN by
 \citet{Padovani04}.  A similar proportion (40\%) of all 142
 radio-faint \emph{Chandra} sources have low, measured photon indices
 characteristic of obscured AGN.

The stacked-average photon index for the whole
\emph{Chandra} HDF(N) field becomes flatter for lower X-ray count
rates and then slightly steepens again 
\citep{Alexander03XIII}.  The composite X-ray spectra of the most
obscured \emph{Chandra} sources ($N_{\mathrm{H}} > 5\times10^{27}$
m$^{-2}$) show an upturn at rest frame energies $<4$ keV, attributed
to star formation \citep{Alexander05a}.  This is independent evidence
for the association between obscured AGN and starburst activity.

X-ray sources associated with individual galaxies in the 8-arcmin
field are smaller than the $\approx1$\arcsec\/ resolution of
\emph{Chandra}, and none are among the extended emission sources
identified by
\citet{Bauer02a}.  About 20\% (29/142) of the radio-faint X-ray
sources have $F_{\mathrm{X}}>0.3\times10^{-18}$ and $\Gamma>1.4$.
Equation~\ref{Bauer} predicts a radio flux density $>40$ $\mu$Jy from
such sources if the X-ray emission is of starburst origin, so they
would have been detected as radio-bright if the radio and X-ray
emission had a common starburst origin. Instead, it is likely that these
radio-quiet sources  contain unobscured X-ray AGN.

Almost all
the remaining 40\% of the  radio-faint sources, with
$F_{\mathrm{X}}\le0.3\times10^{-18}$, do not have measured photon
indices
\citep{Bauer04}.  \citet{Reddy04} stacked \emph{Chandra} soft-band and
VLA-only radio HDF(N) flux densities for (rest-frame) UV-selected
objects at a mean redshift of $\sim2$. Their measurements were
consistent with the relationship found by \citet{Bauer02}, giving an
average SFR of $\sim50$ M$_{\odot}$ yr$^{-1}$. Stacking \emph{Chandra}
data for Lyman Break galaxies out to $z=6$, gave similar results.  Their
selection criteria (optical detection, soft-band dominated) excluded
objects like the SMGs and the most vigorous radio starbursts
associated with type-II AGN, so these sources are probably analogues
of the local ULIRGs.

In Section~\ref{z} we used the Kolmogorov-Smirnov test to show that
there seemed to be a change in the relationship between radio and
X-ray sources at redshifts above and below $\sim1.1$.  We also found
(Section~\ref{empire}) a starburst-like relationship between
$L_{\mathrm{R}}$ and $L_{\mathrm{X}}$ for most low-redshift sources,
but only very weak relationships at high redshift, most significant
for the radio starbursts hosting type-II AGN.  The correlation between
radio-faint radio flux density and harder X-ray emission seen in
Fig.~\ref{BINBYCXOFLUX.PS} (upper) is consistent with a continuation
of this relationship for the radio-faint sources. The correlation with
harder photon index is also consistent with type-II AGN powering the
the dominant population of distant X-ray sources with radio
counterparts of a few tens of $\mu$Jy.

In summary, we predict that future instruments will confirm that the
 association between the presence of X-ray and radio emission extends
 to radio flux densities $\la10$ $\mu$Jy arcsec$^{-2}$. The available
 evidence suggests that two fifths of these sources will be found to
 be radio starbursts hosting obscured X-ray AGN, probably at high
 redshifts, one fifth are radio-quiet, unobscured AGN and the
 remainder are probably ULIRGs.

\section{Conclusions}
\label{conclusions}
We have used Virtual Observatory tools and  RadioNet software
to compare the radio and X-ray properties of
galaxies at $z\la4.5$ in the HDF(N).  92 radio-bright objects ($S_{\mathrm{R}}>40$ $\mu$Jy) are resolved by
MERLIN+VLA at 1.4 GHz. 55 of these sources are also securely detected
by \emph{Chandra} and the majority of these sources have
spectral/photon indices based on measurements at more than one
frequency. Where spectral/photon indices are estimated, we have not used
the values as primary diagnostics.

\subsection{The majority of radio starbursts in the HDF(N) have X-ray counterparts}

The combination of radio morphologies and spectral indices, in some
cases supported by rest-frame MIR measurements, provides reliable
diagnostics for the origins of the radio emission from most of the
sources (Section~\ref{sec:class}; \citealt{Muxlow05}).  70\% of radio
starbursts have X-ray counterparts.  Radio starbursts outnumber radio
AGN by 3:1 for classified sources in the whole radio-bright sample and
the same proportion is seen for the 55 sources with X-ray
counterparts.  Seven unresolved 8.4-GHz sources which are not
radio-bright at 1.4 GHz also have X-ray counterparts are not included
in most of this analysis, but they appear to contain a similar
majority of starbursts.  In contrast, the X-ray emission is of AGN
origin in about 80\% of all \emph{Chandra} sources \citep{Bauer02}.

Optical spectroscopy finds approximately equal numbers of star formers
and AGN \citep{Cowie04a} and only 6 optically-classified starbursts in the
HDF(N) have X-ray counterparts (along with 29 optical AGN). Similarly,
only  a few percent of the starbursts found using
$BzK$ criteria,
\citep{Daddi04a}, have
soft-X-ray counterparts \citep{Daddi05} (hard-X-ray sources were
excluded from their sample).  These results imply that many radio
starbursts which are too obscured to be identified using optical or
even NIR spectroscopy and photometry, are a separate population with a
much higher proportion of X-ray counterparts.
 
\subsection{High-luminosity radio and X-ray emission has separate
  origins with the same galaxies}

There is no discernible relationship between  either
the radio and X-ray flux densities or the $K$-corrected luminosities
for the cross-matched sample as a whole.
The close relationship between $L_{\mathrm{R}}$ and
  $L_{\mathrm{X}}$ established by \citet{Bauer02} breaks down at
  $z\ga1.3$, even if only radio starbursts are considered
  (Section~\ref{empire}).  The X-ray emission predicted from the radio
luminosity of these sources is on average less than 1/3 of the
observed X-ray luminosity and even after subtracting the potential
starburst contribution, the hard-band X-ray luminosity of the type-II AGN
still exceeds $10^{35}$ W (Section~\ref{XH}). At least half of the radio
sources with X-ray counterparts have a largest angular size greater
than the \emph{Chandra} resolution of  $\approx1$\arcsec\/  but all the X-ray
sources are unresolved.

The presence of detectable radio emission is significantly correlated
with the probability of also detecting X-rays, and vice versa \citep{Bauer02}
but our results strongly suggest that  the radio
and X-ray emission originates from separate phenomena in most sources at
$z\ga1.1$.

\subsection{A distinct population of high-redshift radio/sub-mm
  starbursts associated with X-ray-selected obscured AGN}

The fraction of X-ray sources with radio counterparts increases
significantly at $z\ga1.1$ (Fig.~\ref{ZX_AGN2_XRAGN2.eps}).
Section~\ref{sbagn} sums up the evidence that high-$z$ radio- or
sub-mm-selected starbursts are a separate population from starforming
galaxies at $z\la1.3$ and indeed exceed the activity seen in local
ULIRGs (e.g. Section~\ref{scuba} and references therein.)

A hard photon index $\Gamma\le1$ (at rest-frame 0.5--8 keV) combined
with $L_{\mathrm{X}}>10^{35}$ W shows the presence of an obscured type-II AGN
and 18 such objects identified by \citet{Padovani04} have radio
counterparts in the HDF(N).  Model X-ray spectra extending to high
rest-frame energies \citep{Alexander05a} suggest that type-II AGN at $z\ga2$
could have slightly higher $\Gamma$ as measured by \emph{Chandra}.
This is supported by the $L_{\mathrm{R}}$--$L_{\mathrm{X}}$
relationship and $\Gamma$--$z$ distribution of type-II AGN with radio
counterparts (Fig.~\ref{GZ.PS}) which leads us to propose that
J123642+621331, at $z=4.424$ with observed
$\Gamma=1.35^{+0.34}_{-0.40}$, also contains a type-II AGN. Twelve of the
type-II AGN (including this source) have radio starburst hosts.

The great majority (22/27) of 15 or 16~$\mu$m detections among the
radio+X-ray sources are not type-II AGN hosts. They have a mean
redshift of 1 and a mean photon index of 1.6.  They appear more
analogous to local ULIRGS than to the more extreme SMGs.  Well-studied
ULIRGs such as Arp 220 or Markarian 231 possess nuclear starbursts
which would be barely resolved at the maximum detectable $z\sim1$.
\citet{Bauer04} noted that the relative number counts of starburst and
AGN X-ray sources changed from a large AGN majority over most of the
flux density range sampled by \emph{Chandra} to a higher proportion of
starbursts among the faintest sources, likely to correspond to the
less active tail of a relatively nearby population. The faintest $\sim40\%$ of
X-ray sources which are radio-quiet, but have a significant radio flux
density revealed by stacking, (Section~\ref{rfn}), may also be ULIRGs.

 In contrast, the SMG counterparts have a mean redshift of 1.8 and an
average photon index of 0.6, suggesting that type-II AGN contribute
substantially more of the X-ray emission. \citet{Alexander05b} suggest
that between a third and a half of the entire SCUBA source population
contain AGN. 2/3 of radio-bright sources with X-ray and SCUBA emission
contain type-II AGN. About $40\%$ of the radio-faint sources found by
stacking are associated with hard X-ray sources are probably also
radio starbursts hosting obscured AGN.  The radio starbursts appear to
be extended over 2 -- 10 kpc or more, an order of magnitude larger
than those in the local universe.  The SMG (at a median redshift of at
least 2.2) have star formation rates of the order of 1000 M$_{\odot}$
yr$^{-1}$, also about 10 times higher than that of local ULIRGS.

\subsection{Implications for galaxy evolution}

Both starburst activity and feeding a black hole probably result from
major mergers, which are increasingly common at $z > 1.5$ and dominate
galaxy growth at $z > 2$ \citep{Conselice05}.  The comoving luminosity
densities for AGN and for starbursts increases as $(1+z)^{\ga3}$ for
$z\la1.5$, using optical, IR, X-ray and radio classifications and
luminosities \citep{Barger05, Cowie04a}.  Current estimates of the
star formation rate as a function of redshift \citep{Hopkins06}
suggest that the star formation rate continues to increase out to
redshifts of 3 or more. It is possible that sub-mm galaxies at unknown
redshifts may be even more vigorous starbursts at $z\ga3$
\citep{Ivison07}.

 The one radio-bright object in the HDF(N) with a known redshift greater than 3,
J123642+621331 at $z=4.424$, appears to contain both a (probably
obscured) AGN and a very productive starburst. Stacking 1.4-GHz
emission from radio-faint sources at the position of objects detected
by \emph{Chandra} shows a statistically significant excess which rises
with decreasing $\Gamma$ and with increasing hard-band X-ray flux
densities (Section~\ref{rfn}). Eleven individual radio-faint
counterparts to X-ray sources are $>5\sigma$ ($>20$~$\mu$Jy), all of
which appear extended and lacking radio hotspots, consistent with
these sources being a faint or high-redshift tail of the radio
starbursts associated with type-II AGN.

 Within the next 2 years, \emph{e}-MERLIN+EVLA images should reveal
many more high-redshift $\mu$Jy galaxies with optical, \emph{Spitzer}
or SCUBA(2) counterparts, pushing measurements of the star formation
rate (and redshifts) back to $z\ga5$, whilst the ongoing increase in
capacity of the VLBI correlator at JIVE will enable distant compact
radio AGN cores to be distinguished from even sub-kpc-scale starbursts
\citep{Garrett05}.

\begin{acknowledgements}

This work is based in part on observations made with the Multi Element
Radio-Linked Interferometer Network (MERLIN), operated by Manchester
University on behalf of PPARC, and with the Very Large Array, a
facility of the NSF operated under cooperative agreement by Associated
Universities, Inc. This work has benefited from research funding from
the EC Framework 6 Programme under RadioNet R113CT 2003 5058187. We
employed software provided by the UK AstroGrid Virtual Observatory
Project, funded by PPARC and the Framework 6 Programme and we also
used the VizieR catalogue access tool and the ADS bibliographic
service.

We thank the MERLIN and VLA staff for considerable assistance with
observations.  AMSR acknowledges the hospitality provided by CDS,
where she was a `professeur invit\'{e}' during part of this work. We
thank P. Padovani (ESO) for helpful discussions and we are extremely
grateful to the anonymous referee for improving the accuracy, clarity
and consistency of the paper.

\end{acknowledgements}

\bibliographystyle{aa}

\bibliography{RichardsAMS_AA7598.bib}

\begin{sidewaystable*}
\begin{minipage}[t][180mm]{\textwidth}
\caption{\label{tab1} Radio and X-ray sources detected in both
regimes.  The first 55 were selected at 1.4 GHz \protect{\citep{Muxlow05}}.  The last 7 were selected at 8.4 GHz  \protect{\citep{Richards98}}.
Columns (1)-(8) give the name, position, flux density (at the
selection frequency), spectral index and uncertainties of radio
sources. Columns (9)-(11) give the redshift, its uncertainty and the
reference (see footnote). Column (12) gives the separation between the
radio and X-ray peaks. Columns (13)-(17) give the number, position,
flux density, photon index and uncertainties of X-ray sources \protect{\citep{Alexander03XIII}}.  Numbers in {\em italics} signify values of
$\alpha$ or $\Gamma$ estimated from single measurements, and
photometric redshifts.  Columns (18) indicates whether the object was detected by {\em ISO} (I), {\em Spitzer} (S) or SCUBA
(SMM). See text for further details of crossmatching and estimating
uncertainties not given in the original papers.} \centering
\begin{tabular}{cccrrrrrlllrrrrllc}
\hline\hline
Name& RA& Dec& $S_{\mathrm{R1.4}}$&LAS& $\alpha$&$\sigma_{\alpha_-}$&$\sigma_{\alpha_+}$ &$z$&$\sigma_z$& Ref.&Sep.& ABB&$F_{\rm x}$&$\Gamma$&$\sigma_{\Gamma_-}$& $\sigma_{\Gamma_+}$&IR/SMM\\
 & (J2000)&(J2000)&($\mu$Jy)&(\arcsec)& & & & & &&(\arcsec)&\multicolumn{3}{c}{($10^{-18}$ W m$^{-2}$)}&&&\\
(1)&(2)&(3)&(4)&(5)&(6)&(7)&(8)&(9)&(10)&(11)&(12)&(13)&(14)&(15)&(16)&(17)&(18)\\
\hline
&&&&&&&&&&&&&&&&&\\
J123606+620951	& 12 36 06.6128	& +62 09 51.141	& 196	& 0.9	&$>$0.56& 0.07	& 1.14	& 0.6379& 0.003& CBHCS04& 0.31	& 76 	& 3.22	&$<$--0.91	& 0.16	& 0.09	& --	\\
J123606+621021	& 12 36 06.8493	& +62 10 21.437	&  74	& 0.7	& {\em 0.80}& 0.50	& 0.90	& 2.51	& 0.003& CSWMI04& 0.08	& 79 	& 0.74	&$<$0.12	& 0.37	& 0.6	&SMM\\
J123608+621035	& 12 36 08.1195	& +62 10 35.898	& 217	& 0.6	& 0.36	& 0.08	& 0.08	& 0.681	& 0.003& CBHCS04& 0.10	& 82 	& 1.89	& 0.18		& 0.21	& 0.20	&--	\\
J123608+621553	& 12 36 08.2421	& +62 15 53.094	&  59	& 0.5	&{\em --0.24}& 0.10& 0.90	& 0.4593& 0.003& CBHCS04& 0.15	& 83 	& 3.57	& --0.32	& 0.18	& 0.18	& S	\\
J123612+621138	& 12 36 12.0272	& +62 11 38.733	&  75	& 1.2	& {\em 0.80} & 0.50	& 0.90	& 0.275	& 0.003& CBHCS04& 0.14	& 93 	&17.71	& 1.41		& 0.04	& 0.04	&--	\\
&&&&&&&&&&&&&&&&&\\												 		 				
J123616+621513	& 12 36 16.1419	& +62 15 13.937	&  53	& 1.0	& {\em 0.80} & 0.50	& 0.90	& 2.58	& 0.003& CSWMI04& 0.33	& 109	& 1.02	& 0.97		& 0.21	& 0.19	& SMM\\
J123617+621011	& 12 36 17.0801	& +62 10 11.306	&  62	& 0.6	& {\em 0.40} & 0.50	& 1.30	& 0.845	& 0.003& H+01   & 0.33	& 113	& 4.95	& 1.55		& 0.08	& 0.08	&--	\\
J123618+621635	& 12 36 18.0170	& +62 16 35.270	&  47	& 2.5	& {\em 0.00} & 0.50	& 1.40	& 0.679	& 0.003& H+01	 & 0.29	& 115	&15.64	& 1.45		& 0.04	& 0.04	&--	\\
J123619+621252	& 12 36 19.4784	& +62 12 52.581	& 108	& 0.4	&$>$0.80& 0.09	& 0.90	& 0.473	& 0.003& CBHCS04& 0.27	& 119	& 0.24	& {\em 1.4}		& 0.6	& 0.6	&--	\\
J123621+621109	& 12 36 21.2256	& +62 11 09.007	&  52	& 3.2	&$>$0.86& 0.17	& 0.84	& 1.014	& 0.003& C+00   & 0.23	& 127	& 3.86	& --0.73	& 0.21	& 0.21	&--	\\
&&&&&&&&&&&&&&&&&\\												 		 				
J123622+621544	& 12 36 22.4753	& +62 15 44.776	&  83	& 3.1	&$>$0.60& 0.11	& 0.80	& 0.6471& 0.003& CBHCS04& 0.57	& 132	& 0.32	& 1.06		& 0.48	& 0.38	& S	\\
J123622+621629	& 12 36 22.6535	& +62 16 29.718	&  70	& 1.3	& {\em 0.80} & 0.50	& 0.90	& 2.4	& 0.003& CSWMI04& 0.09	& 135	& 1.02	&$<$--0.5	& 0.25	& 0.5	& SMM\\
J123624+621743	& 12 36 24.7685	& +62 17 43.160	&  78	& 0.9	& {\em 0.40} & 0.80	& 1.30	& --	&-- 	&--	 & 0.55	& 143	& 0.47	&$<$0.29	& 0.46	& 0.6	&--	\\
J123629+621046	& 12 36 29.1240	& +62 10 45.984	&  81	& 2.7	&$>$0.86& 0.11	& 0.84	& 1.013	& 0.003& C+00   & 0.13	& 158	& 2.31	& 0.18		& 0.16	& 0.15	& S,SMM\\
J123630+620923	& 12 36 30.0516	& +62 09 23.895	&  46	& 0.6	& {\em 0.80} & 0.50	& 0.90	& 0.953	& 0.003& B+02   & 0.20	& 164	& 1.93	&$<$--0.77	& 0.19&0.23	&--	\\
&&&&&&&&&&&&&&&&&\\												 		 				
J123631+620957	& 12 36 31.2450	& +62 09 57.791	& 152	& 0.8	&$>$0.99& 0.07	& 0.71	& --	& --	&--	 & 0.49	& 167	& 0.21	& {\em 1.4}		& 0.6	& 0.6	&--	\\
J123632+620759	& 12 36 32.5583	& +62 07 59.846	&  90	& 2.6	&$>$0.58& 0.11	& 1.12	& 1.9939& 0.003& CBHCS04& 0.23	& 171	& 1.83	& 0.05		& 0.23	& 0.22	& SMM\\
J123633+621005	& 12 36 33.7269	& +62 10 05.962	&  46	& 1.2	&$>$0.98& 0.19	& 0.72& 1.016	& 0.003& CBHCS04& 0.48	& 177	& 0.86	& 1.42		& 0.22	& 0.19	&S	\\
J123634+621213	& 12 36 34.4701	& +62 12 13.006	& 233	& 1.2	& 0.74	& 0.06	& 0.06	& 0.456	& 0.003& CBHCS04& 0.07	& 180	& 0.44	& 1.96		& 0.32	& 0.28	& I,S\\
J123634+621241	& 12 36 34.5168	& +62 12 41.107	& 230	& 1.0	& 0.74	& 0.06	& 0.06	& 1.219	& 0.003& C+00   & 0.15	& 182	& 0.29	&$>$1.63	& 0.37	& 0.28	& I,S,SMM\\
&&&&&&&&&&&&&&&&&\\												 		 				
J123635+621424	& 12 36 35.5839	& +62 14 24.049	&  87	& 0.3	&$>$0.87& 0.10	& 0.83	& 2.011	& 0.003& D+01 	 & 0.06	& 190	& 2.52	& 0.25		& 0.14  &0.13	& I,S,SMM\\
J123636+621320	& 12 36 36.9061	& +62 13 20.337	&  50	& 0.7	& {\em 0.80} & 0.50	& 0.90	& --	& --	&--	 & 0.34	& 196	& 0.41	& 0.1		& 0.48	& 0.47	&--	\\
J123641+620948	& 12 36 41.5511	& +62 09 48.232	&  75	& 0.6	&$>$0.56& 0.12	& 1.14	& 0.518	& 0.003& C+00   & 0.15	& 217	& 1.48	& --0.65	& 0.38	& 0.43	& S	\\
J123642+621331	& 12 36 42.0916	& +62 13 31.426	& 467	& 0.4	& 0.94	& 0.06	& 0.06	& 4.424	& 0.003& W+99	 & 0.22	& 220	& 0.28	& 1.35		& 0.40	& 0.34	& I	\\
J123642+621545	& 12 36 42.2123	& +62 15 45.521	& 150	& 1.7	& 0.50	& 0.07	& 0.07	& 0.857	& 0.003& H+01	 & 0.09	& 222	& 2.46	& 1.22		& 0.14	& 0.13	& I	\\
&&&&&&&&&&&&&&&&&\\												 		 				
J123644+621133	& 12 36 44.3870	& +62 11 33.145	&1290	& 12.0	& 0.30	& 0.05	& 0.05	& 1.05	& 0.003& C+00   & 0.18	& 230	& 0.24	& {\em 1.4}		& 0.6	& 0.6	& I	\\
J123645+620754	& 12 36 45.8620	& +62 07 54.190	&  48	& 3.0	& {\em 0.40} & 0.80	& 1.30	& 1.433	& 0.003& CBHCS04& 0.47	& 237	& 0.34	&$>$1.1		& 0.9	& 0.37	&--	\\
J123646+621448	& 12 36 46.0629	& +62 14 48.713	& 124	& 0.5	& 0.84	& 0.12	& 0.86	& --	& --	&--	 & 0.19	& 239	& 0.08	& {\em 1.4}		& 0.6	& 0.6	&--	\\
J123646+621404	& 12 36 46.3321	& +62 14 04.693	& 179	& 0.5	& --0.04& 0.06	& 0.06	& 0.961	& 0.003& CBHCS04& 0.02	& 240	&24.63	& 0.67		& 0.04	& 0.04	& I,S\\
J123646+621445	& 12 36 46.7360	& +62 14 45.640	& 117	& 2.3	& 0.98	& 0.15	& 0.15	& --	& --	&--	 & 0.28	& 243	& 0.09	& {\em 1.4}		& 0.6	& 0.6	&--	\\
\end{tabular}
\vfill
\end{minipage}
\end{sidewaystable*}

\begin{sidewaystable*}
\begin{minipage}[t][180mm]{\textwidth}
\addtocounter{table}{-1}
\caption{ continued.}
\begin{tabular}{cccrrrrrlllrrrrllcc}
\hline\hline
Name& RA& Dec& $S_{\mathrm{R1.4}}$&LAS& $\alpha$&$\sigma_{\alpha_-}$&$\sigma_{\alpha_+}$ &$z$&$\sigma_z$& Ref.&Sep.& ABB&$F_{\rm x}$&$\Gamma$&$\sigma_{\Gamma_-}$& $\sigma_{\Gamma_+}$&IR/SMM\\
 & (J2000)&(J2000)&($\mu$Jy)&(\arcsec)& & & & & & &(\arcsec)&\multicolumn{3}{c}{($10^{-18}$ W m$^{-2}$)}&&&\\
(1)&(2)&(3)&(4)&(5)&(6)&(7)&(8)&(9)&(10)&(11)&(12)&(13)&(14)&(15)&(16)&(17)&(18)\\
\hline
&&&&&&&&&&&&&&&&&\\
J123646+620833	& 12 36 46.6587	& +62 08 33.291	&  80	& 0.8	&$>$0.61& 0.11	& 1.09	& 0.9712& 0.003& CBHCS04& 0.59	& 244	& 0.17	& {\em 1.4}		& 0.6	& 0.6	&--	\\
J123649+621313	& 12 36 49.7432	& +62 13 13.065	&  49	& 1.0	& 0.72	& 0.15	& 0.15	& 0.475	& 0.003& C+00   & 0.27	& 260	& 0.16	& {\em 1.4}		& 0.6	& 0.6	& I,S,SMM\\
J123651+621030	& 12 36 51.1223	& +62 10 30.955	&  95	& 1.2	& 0.74	& 0.17	& 0.17	& 0.41	& 0.003& C+00   & 0.59	& 265	& 0.21	&$>$1.61	& 0.39	& 0.31	& I,S	\\
J123651+621221	& 12 36 51.7258	& +62 12 21.435	&  49	& 1.2	& 0.71	& 0.12	& 0.12	& {\em 2.71}	& 0.2	& CBDP03  & 0.06	& 267	& 2.94	& 0.56		& 0.12	& 0.12	& I,S\\
J123652+621444	& 12 36 52.8839	& +62 14 44.076	& 168	& 1.3	& --0.12& 0.07	& 0.07	& 0.321	& 0.003& CBHCS04& 0.05	& 274	& 0.46	& 2.1		& 0.33	& 0.29	&--	\\
&&&&&&&&&&&&&&&&&\\												 		 				
J123653+621139	& 12 36 53.3629	& +62 11 39.647	&  65	& 0.3	& 0.77	& 0.12	& 0.12	& 1.275	& 0.003& C+00   & 0.07	& 277	& 0.21	&$>$1.93	& 0.07	& 0.26	& I	\\
J123655+620808	& 12 36 55.9397	& +62 08 08.163	& 106	& 0.7	&$>$0.85& 0.09	& 0.85	& 0.792	& 0.003& CBHCS04& 0.66	& 288	& 0.25	&$>$1.35	& 0.65	& 0.32	& S	\\
J123656+621301	& 12 36 56.9170	& +62 13 01.783	&  49	& 1.2	&$>$1.22& 0.16	& 0.48	& 0.474	& 0.003& C+00   & 0.18	& 294	& 0.76	& 1.8		& 0.23	& 0.21	& I	\\
J123659+621449	& 12 36 59.9150	& +62 14 49.503	&  47	& 1.0	& {\em 0.80} & 0.50	& 0.90	& 0.761	& 0.003& CBHCS04& 0.30	& 310	& 0.11	& {\em 1.4}		& 0.6	& 0.6	& I	\\
J123701+621146	& 12 37 01.5745	& +62 11 46.814	& 128	& 3.0	& 0.67	& 0.08	& 0.08	& {\em 1.52}& 0.38	& BCB+02 & 0.69	& 317	& 0.09	& {\em 1.4}		& 0.6	& 0.6	& I,SMM\\
&&&&&&&&&&&&&&&&&\\												 		 				
J123704+620755	& 12 37 04.1120	& +62 07 55.484	&  50	& 1.0	& {\em 0.80} & 0.50	& 0.90	& 1.253	& 0.003& BCB+02 & 0.18	& 331	& 6.34	& 1.83		& 0.08	& 0.07	&--	\\
J123705+621153	& 12 37 05.8599	& +62 11 53.541	&  52	& 1.5	&$>$1.27& 0.17	& 0.43	& 0.902	& 0.003& CBHCS04& 0.25	& 337	& 0.12	& {\em 1.4}		& 0.6	& 0.6	& I	\\
J123707+621408	& 12 37 07.2209	& +62 14 08.208	&  45	& 0.9	& 0.29	& 0.16	& 0.16	& 2.48	& 0.003& CSWMI04& 0.34	& 347	& 0.98	& 0.39		& 0.25	& 0.23	& SMM\\
J123708+621056	& 12 37 08.3663	& +62 10 56.049	&  50	& 0.7	& 0.35	& 0.17	& 0.17	& 0.422	& 0.003& CBHCS04& 0.29	& 353	& 0.28	&$>$1.46	& 0.54	& 0.30	&S	\\
J123709+620837	& 12 37 09.4315	& +62 08 37.580	&  45	& 0.5	& --0.35& 0.11	& 0.11	& 0.907	& 0.003& CBHCS04& 0.32	& 358	& 0.87	& 1.08		& 0.31	& 0.28	&--	\\
&&&&&&&&&&&&&&&&&\\												 		 				
J123709+620841	& 12 37 09.7518	& +62 08 41.249	&  72	& 0.8	& {\em 0.00} & 0.40	& 0.12	& 0.902	& 0.003& CBHCS04& 0.52	& 359	& 0.30	&$>$1.09	& 0.91	& 0.43	& S	\\
J123711+621330	& 12 37 11.2549	& +62 13 30.846	&  67	& 1.2	& 0.69	& 0.13	& 0.13	&{\em 1.112}& 0.15	& B+02   & 0.88	& 366	& 0.56	& 0.62		& 0.31	& 0.28	&--	\\
J123711+621325	& 12 37 11.9865	& +62 13 25.771	& 132	& 0.8	&$>$1.16& 0.08	& 0.54	& 1.99	& 0.003& CSWMI04& 0.38	& 368	& 0.92	&$<$--0.44	& 0.26	& 0.56	& SMM\\
J123716+621512	& 12 37 16.3740	& +62 15 12.343	& 187	& 0.4	& 0.41	& 0.09	& 0.09	& 0.23	& 0.003& CBHCS04& 0.31	& 388	& 0.23	&$>$1.52	& 0.48	& 0.32	&--	\\
J123716+621643	& 12 37 16.5410	& +62 16 43.799	&  66	& 0.9	& {\em 0.80} & 0.50	& 0.90	& 0.557	& 0.003& C+00   & 0.36	& 389	& 0.42	& 0.51		& 0.47	& 0.41	&--	\\
&&&&&&&&&&&&&&&&&\\												 		 				
J123716+621733	& 12 37 16.6811	& +62 17 33.327	& 346	& 0.6	&$>$0.76& 0.06	& 0.64	& 1.146	& 0.003& CBHCS04& 0.22	& 390	&21.69	& 1.03		& 0.04	& 0.04	&--	\\
J123716+621007	& 12 37 16.8252	& +62 10 07.401	&  63	& 0.7	& {\em 0.80} & 0.50	& 0.90	& 0.411	& 0.003& C+00   & 0.50	& 392	& 0.13	& {\em 1.4}		& 0.6	& 0.6	&--	\\
J123721+621129	& 12 37 21.2539	& +62 11 29.954	& 382	& 0.8	& --0.28 & 0.06	& 0.06	& {\em 1.56}& 0.39	& C+00	 & 0.30	& 403	& 0.51	& 0.28		& 0.43	& 0.40	&--	\\
J123725+620856	& 12 37 25.0112	& +62 08 56.374	&  90	& 0.4	& {\em 0.40} & 0.80	& 1.30	&{\em 0.984}& 0.15	& B+02   & 0.33	& 416	& 3.57	& 0.85		& 0.13	& 0.12	&--	\\
J123734+620931	& 12 37 34.2420	& +62 09 31.930	& 142	& 2.8	& {\em 0.40} & 0.80	& 1.30	& 0.189 & 0.003& BCB+02 & 0.24	& 436	& 0.75	&$>$1.17	& 0.83	& 0.36	&--	\\
&&&&&&&&&&&&&&&&&\\
\hline
Name& RA& Dec& $S_{\rm R8.4}$&LAS&
$\alpha$&$\sigma_{\alpha_-}$&$\sigma_{\alpha_+}$ &$z$&$\sigma_z$& Ref.&Sep.& ABB&$F_{\rm x}$&$\Gamma$&$\sigma_{\Gamma_-}$& $\sigma_{\Gamma_+}$&IR/SMM\\
 & (J2000)&(J2000)&($\mu$Jy)&(\arcsec)& & & & & &&(\arcsec)&\multicolumn{3}{c}{($10^{-18}$ W m$^{-2}$)}&&&\\
\hline
&&&&&&&&&&&&&&&&&\\
J123637+621135	& 12 36 37.002	& +62 11 35.14  &17.5& -- & 0.6	& 0.2	& 0.2	& 0.078	& 0.003&CBHCS04& 1.25	& 197	& 0.14	&{\em 1.4}		& 0.6	& 0.6	& I	\\
J123639+621249	& 12 36 39.881	& +62 12 49.93  &9.8& -- & 1.0	& 0.3	& 0.3	& 0.846	& 0.003& C+00& 0.27	& 211	& 0.22	& 1.3		& 0.4	& 0.3	& I,S	\\
J123644+621249	& 12 36 44.017	& +62 12 49.98  &10.2& -- & 0.7	& 0.2	& 0.2	& 0.556	& 0.003& C+96& 0.18	& 227	& 0.09	&{\em 1.4}		& 0.6	& 0.6	& I,S \\
J123648+621427	& 12 36 48.377	& +62 14 27.65  &9.8& -- & 0.7	& 0.3	& 0.3	& 0.139	& 0.003&CBHCS04& 1.24	& 251	& 0.22	&{\em 1.4}		& 0.6	& 0.6	& I,S	\\
J123652+621354	& 12 36 52.764	& +62 13 54.77  &7.8& -- & $<$0.4	& 0.8	& 0.3	& 1.355	& 0.003&CBHCS04 & 0.69	& 272	& 0.13	&{\em 1.4}		& 0.6	& 0.6	& SMM\\
&&&&&&&&&&&&&&&&&\\																	
J123655+621311	& 12 36 55.402	& +62 13 11.41  &12.3& -- & $<$0.3	& 0.7	& 0.3	& 0.955	& 0.003&CBHCS04& 0.39	& 286	& 1.23	& 1.5		& 0.1	& 0.1	&--	\\
J123658+621434	& 12 36 58.720	& +62 14 34.61  &11.4& -- & $<$0.4	& 0.8	& 0.3	& 0.678	& 0.003&CBHCS04& 0.96	& 304	& 9.18	& 1.6		& 0.05	& 0.05	&--	\\
\end{tabular}
\vfill
\end{minipage}
\end{sidewaystable*}
\newpage
\begin{table*}
\begin{tabular}{ll}
\multicolumn{2}{l}{{\bf Table 1.} continued: References}\\
B+02&\protect{\citet{Bauer02}}\\
BCB+02&\protect{\citet{Barger02}}\\
C+96&\protect{\citet{Cohen96}}\\
C+00&\protect{\citet{Cohen01}}\\
CBDP03&\protect{\citet{Conselice03}}\\
CBHCS04&\protect{\citet{Cowie04b}}\\
CSWMI04&\protect{\citet{Chapman04a}}\\
D+01&\protect{\citet{Dawson01}}\\
H+01 &\protect{\citet{Hornschemeier01}}\\
W+99&\protect{\citet{Waddington99}}\\
\hline
\end{tabular}
\end{table*}

\begin{table*}
\centering
\caption{\label{tab2}Luminosity and classification of cross-matched
  radio and X-ray sources.  Columns (1) and (2) give the radio name
  and rest-frame 1.4-GHz luminosity, columns (3) and (4) give the
  lower and upper luminosity bounds and column (5) gives the
  radio-based classification where SB = starburst, AGN = AGN and U =
  unclassified.  Columns (6) and (7) give the source number from
  Alexander et al. (2003) and the 0.5-8 keV X-ray luminosity, columns
  (8) and (9) give the lower and upper luminosity bounds and column
  (10) indicates whether the hard-band X-ray luminosity is $>10^{35}$ W, indicating an X-ray AGN (AGN-X), or is also classified as an X-ray selected
  type-II AGN (AGN-II) listed by Padovani et al. (2005).  The first 55 sources were selected at 1.4 GHz and the last 7 at 8.4 GHz, references as in Table~\protect{\ref{tab1}}.}
\begin{supertabular}{cccccrcccc}
\hline
\hline
   Name  & $L_{\rm R}$ & $L_{\rm Rmin}$  & $L_{\rm Rmax}$ &Class& ABB & $L_{\rm X}$&$L_{\rm Xmin}$&$L_{\rm Xmax}$& X-ray AGN? \\
& W Hz$^{-1}$& W Hz$^{-1}$& W Hz$^{-1}$ & (radio) & &W&W&W&\\
(1)&(2)&(3)&(4)&(5)&(6)&(7)&(8)&(9)&(10)\\
\hline
	&&&&&&&&&       		 	  	   	   	  	   	            \\
J123606+620951&  2.45E23& 2.27E23& 3.83E23&  U    &76    &1.19E35 &1.05E35 &1.35E35&  AGN-II   \\
J123606+621021&  4.04E24& 1.45E24& 8.64E24&  SB  &79    &4.91E35 &1.07E35 &7.42E35&  AGN-II   \\
J123608+621035&  2.84E23& 2.63E23& 3.05E23&  AGN  &82    &1.34E35 &1.14E35 &1.55E35&  AGN-II   \\
J123608+621553&  2.58E22& 2.21E22& 3.53E22&  SB  &83    &1.04E35 &9.33E34 &1.14E35&  AGN-II   \\
J123612+621138&  1.53E22& 1.24E22& 1.94E22&  SB  &93    &3.29E35 &3.19E35 &3.40E35& AGN-X   \\
	&&&&&&&&&       		 	  	   	   	  	   	            \\
J123616+621513&  3.14E24& 1.08E24& 6.79E24&  SB  &109   &2.10E36 &1.54E36 &2.71E36&  AGN-II \\
J123617+621011&  1.36E23& 9.09E22& 2.46E23&  U    &113   &1.19E36 &1.11E36 &1.26E36&  AGN-X  \\
J123618+621635&  5.08E22& 3.50E22& 8.87E22&  AGN  &115   &2.13E36 &2.07E36 &2.20E36& AGN-X     \\
J123619+621252&  7.50E22& 6.81E22& 1.01E23&  SB  &119   &1.42E34 &9.79E33 &1.91E34& no    \\
J123621+621109&  2.41E23& 1.94E23& 3.88E23&  SB  &127   &2.92E35 &2.44E35 &3.41E35&  AGN-II   \\
	&&&&&&&&&       		 	  	   	   	  	   	            \\
J123622+621544&  1.09E23& 9.73E22& 1.54E23&  SB  &132   &3.22E34 &2.32E34 &4.29E34& no    \\
J123622+621629&  3.35E24& 1.26E24& 7.06E24&  SB  &135   &2.92E35 &1.07E35 &3.96E35&  AGN-II   \\
J123624+621743&  --     &  --    &   --   &  U    &143   &  --    &   --   &    -- &  AGN-II     \\
J123629+621046&  3.75E23& 3.28E23& 5.99E23&  SB  &158   &3.30E35 &2.87E35 &3.76E35&  AGN-II   \\
J123630+620923&  1.74E23& 1.08E23& 2.84E23&  SB  &164   &1.31E35 &1.05E35 &1.55E35&  AGN-II   \\
	&&&&&&&&&       		 	  	   	   	  	   	            \\
J123631+620957&  --     &  --    &   --   &  U    &167   &  --    &   --   &    -- & no   \\
J123632+620759&  1.97E24& 1.66E24& 4.41E24&  U    &171   &7.50E35 &5.52E35 &9.59E35&  AGN-II   \\
J123633+621005&  2.33E23& 1.82E23& 3.58E23&  SB  &177   &2.95E35 &2.46E35 &3.48E35& AGN-X     \\
J123634+621213&  1.45E23& 1.36E23& 1.54E23&  SB   &180   &2.98E34 &2.50E34 &3.52E34& no    \\
J123634+621241&  1.57E24& 1.46E24& 1.69E24&  SB   &182   &1.82E35 &1.32E35 &2.45E35&  AGN-X    \\
	&&&&&&&&&       		 	  	   	   	  	   	            \\
J123635+621424&  2.69E24& 2.29E24& 5.16E24&  SB+AGN  &190   &1.30E36 &1.09E36 &1.53E36&  AGN-II   \\
J123636+621320&  --     &  --    &   --   &  SB  &196   &  --    &   --   &    -- & no   \\
J123641+620948&  5.82E22& 5.13E22& 8.67E22&  SB &217   &4.57E34 &3.54E34 &5.56E34&  AGN-X    \\
J123642+621331&  1.87E26& 1.65E26& 2.08E26&  SB+AGN   &220   &4.13E36 &1.65E36 &7.05E36& AGN-II?    \\
J123642+621545&  3.62E23& 3.33E23& 3.91E23&  SB+AGN  &222   &4.99E35 &4.46E35 &5.56E35& AGN-X    \\
	&&&&&&&&&       		 	  	   	   	  	   	            \\
J123644+621133&  4.39E24& 4.13E24& 4.65E24&  AGN  &230   &8.78E34 &4.59E34 &1.30E35& no    \\
J123645+620754&  3.73E23& 1.00E23& 8.10E23&  U    &237   &2.02E35 &1.19E35 &3.74E35&   AGN-X   \\
J123646+621448&  --     &  --    &   --   &  U    &239   &  --    &   --   &    -- & no   \\
J123646+621404&  3.94E23& 3.65E23& 4.23E23&  AGN  &240   &4.46E36 &4.31E36 &4.62E36&  AGN-II   \\
J123646+621445&  --     &  --    &   --   &  U    &243   &  --    &   --   &    -- & no   \\
\end{supertabular}
\end{table*}
\begin{table*}
\centering
\addtocounter{table}{-1}
\caption{ continued.}
\begin{supertabular}{cccccrcccc}
\hline
\hline
   Name  & $L_{\rm R}$ & $L_{\rm Rmin}$  & $L_{\rm Rmax}$ &Class& ABB & $L_{\rm X}$&$L_{\rm Xmin}$&$L_{\rm Xmax}$& X-ray AGN? \\
& W Hz$^{-1}$& W Hz$^{-1}$& W Hz$^{-1}$ &(radio) & &W&W&W&\\
(1)&(2)&(3)&(4)&(5)&(6)&(7)&(8)&(9)&(10)\\
\hline

	&&&&&&&&&       		 	  	   	   	  	   	            \\
J123646+620833&  2.80E23& 2.45E23& 4.90E23&  SB  &244   &5.17E34 &2.45E34 &8.03E34&   AGN-X   \\
J123649+621313&  3.33E22& 2.81E22& 3.84E22&  SB   &260   &9.61E33 &6.49E33 &1.30E34& no    \\
J123651+621030&  4.65E22& 4.13E22& 5.17E22&  SB   &265   &9.84E33 &7.54E33 &1.25E34& no    \\
J123651+621221&  2.96E24& 2.06E24& 3.86E24&  SB   &267   &3.94E36 &3.04E36 &4.86E36&  AGN-II   \\
J123652+621444&  3.74E22& 3.48E22& 4.01E22&  AGN  &274   &1.44E34 &1.24E34 &1.66E34& no    \\
	&&&&&&&&&       		 	  	   	   	  	   	            \\
J123653+621139&  5.14E23& 4.33E23& 5.96E23&  SB   &277   &1.89E35 &1.37E35 &2.29E35& AGN-X     \\
J123655+620808&  2.60E23& 2.34E23& 3.91E23&  U    &288   &4.59E34 &3.04E34 &6.84E34& no    \\
J123656+621301&  4.02E22& 3.39E22& 4.97E22&  SB   &294   &5.30E34 &4.64E34 &6.03E34& no    \\
J123659+621449&  1.01E23& 6.91E22& 1.55E23&  SB   &310   &1.89E34 &1.01E34 &2.88E34& no    \\
J123701+621146&  1.48E24& 4.83E23& 2.48E24&  SB   &317   &8.12E34 &7.01E33 &1.58E35&   AGN-X   \\
	&&&&&&&&&       		 	  	   	   	  	   	            \\
J123704+620755&  3.87E23& 2.18E23& 6.77E23&  SB  &331   &5.03E36 &4.70E36 &5.40E36& AGN-X    \\
J123705+621153&  2.32E23& 1.88E23& 3.06E23&  SB   &337   &3.06E34 &1.56E34 &4.69E34& no    \\
J123707+621408&  1.25E24& 9.39E23& 1.57E24&  SB  &347   &8.91E35 &6.14E35 &1.19E36& AGN-II    \\
J123708+621056&  2.28E22& 1.88E22& 2.67E22&  SB  &353   &1.32E34 &1.03E34 &1.71E34& no    \\
J123709+620837&  7.18E22& 5.80E22& 8.55E22&  AGN  &358   &1.83E35 &1.40E35 &2.31E35&  AGN-X   \\
	&&&&&&&&&       		 	  	   	   	  	   	            \\
J123709+620841&  1.42E23& 1.02E23& 1.61E23&  AGN &359   &6.28E34 &3.87E34 &1.05E35&  AGN-X    \\
J123711+621331&  3.49E23& 2.17E23& 4.80E23&  AGN &366   &1.31E35 &8.22E34 &1.82E35&  AGN-II   \\
J123711+621325&  5.44E24& 4.79E24& 8.68E24&  SB  &368   &2.19E35 &8.06E34 &2.93E35&  AGN-II   \\
J123716+621512&  2.38E22& 2.21E22& 2.56E22&  AGN &388   &3.00E33 &2.32E33 &3.80E33& no    \\
J123716+621643&  6.72E22& 5.02E22& 9.52E22&  SB  &389   &2.41E34 &1.72E34 &3.20E34& no    \\
	&&&&&&&&&       		 	  	   	   	  	   	            \\
J123716+621733&  2.05E24& 1.90E24& 3.06E24&  AGN &390   &7.37E36 &7.10E36 &7.64E36&  AGN-II  \\
J123716+621007&  3.16E22& 2.48E22& 4.23E22&  SB  &392   &5.69E33 &3.00E33 &8.73E33& no    \\
J123721+621129&  1.94E24& 9.10E23& 2.97E24&  AGN  &403   &1.71E35 &6.39E34 &2.83E35& AGN-X   \\
J123725+620856&  2.82E23& 9.50E22& 5.54E23&  U    &416   &7.68E35 &5.10E35 &1.02E36&  AGN-X   \\
J123734+620931&  1.20E22& 1.00E22& 1.49E22&  U    &436   &6.10E33 &4.41E33 &8.13E33& no    \\
	&&&&&&&&&       		 	  	   	   	  	   	            \\
J123637+621135&  7.17E20& 1.62E20& 1.62E20&  SB  &197   &1.92E32 &5.75E31 &5.75E31& no\\
J123639+621249&  1.87E23& 7.34E22& 7.34E22&  SB  &211   &4.55E34 &1.24E34 &1.44E34& no\\
J123644+621249&  3.47E22& 2.40E22& 4.53E22&  SB  &227   &7.65E33 &4.40E33 &1.08E34& no\\
J123648+621427&  1.60E21& 5.48E20& 5.48E20&  SB  &251   &9.87E32 &2.13E32 &2.13E32& no\\
J123652+621354&  1.08E23& 8.44E22& 4.89E22&  U    &272   &8.81E34 &5.13E34 &5.13E34& AGN-X\\
	&&&&&&&&&       		 	  	   	   	  	   	            \\
J123655+621311&  5.75E22& 2.85E22& 1.48E22&  AGN  &286   &3.84E35 &4.03E34 &4.03E34& AGN-X\\
J123658+621434&  3.09E22& 1.61E22& 1.08E22&  U    &304   &1.35E36 &5.02E34 &5.02E34& AGN-X \\
\hline
\end{supertabular}
\end{table*}

\end{document}